\documentclass[12pt]{article}
\usepackage[pdftex]{graphicx}

\begin{document}
\title{Statistical and hydrodynamic properties of topological polymers for various graphs showing enhanced short-range correlation}
\author{Erica Uehara and Tetsuo Deguchi}

\maketitle

\begin{center}
Department of Physics, Faculty of Core Research \\ 
Ochanomizu University, \\
2-1-1 Ohtsuka, Bunkyo-ku, Tokyo 112-8610, Japan
\end{center}

\begin{abstract} 
For various polymers with different topological structures we numerically evaluate the mean-square radius of gyration and the hydrodynamic radius systematically through simulation.  We call polymers with nontrivial topology {\it topological polymers}. We evaluate the two quantities both for ideal and real chain models and show that the ratios of the quantities among different topological types do not depend on the existence of excluded volume  if the topological polymers have only up to trivalent vertices, as far as the polymers investigated. We also evaluate the ratio of the gyration radius to the hydrodynamic radius, which we expect to be universal from the viewpoint of renormalization group. Furthermore, we show that the short-distance intrachain correlation is much enhanced for topological polymers expressed with complex graphs. 
\end{abstract}

\newpage 
%
%
\section{Introduction}

Statistical and dynamical properties of polymers with nontrivial topology such as ring polymers  have attracted much interest in several branches of physics, chemistry and biology. For instance, some fundamental properties of ring polymers in solution were studied many years ago；\cite{Kramers,Zimm-Stockmeyer,Casassa,Semlyen} circular DNA have been found in nature in the 1960s and knotted DNA molecules are synthesized in experiments in the 1980s; \cite{Vinograd,Nature-trefoil,DNAknots,Bates}  looped or knotted  proteins have been found in nature during the 2000s.\cite{Taylor,Craik} There are many theoretical studies on knotted ring polymers in solution. \cite{Whittington,Micheletti} Due to novel developments in synthetic chemistry,   polymers with different topological structures are synthesized in experiments during the last decade such as not only ring polymers but also tadpole (or lasso) polymers, double-ring (or di-cyclic) polymers and even complete bipartite graph polymers. \cite{Tezuka2000,Tezuka2001,Grubbs,Takano05,Takano07,Grayson,Tezuka2010,Tezuka2011,Tezuka2014} Lasso polymers are studied also in the dynamics of protein folding. \cite{Sulkowska} We call polymers with nontrivial topology {\it topological polymers}. \cite{Tezuka-book} In order to characterize topological polymers it is fundamental to study the statistical and dynamical properties such as the mean-square radius of gyration and the diffusion coefficient. They are related to experimental results such as the  size exclusion chromatography (SEC) spectrum. 
For multiple-ring (or multi-cyclic) polymers hydrodynamic properties are studied theoretically by a perturbative method.  \cite{Fukatsu-Kurata} For double-ring polymers they are studied by the Monte-Carlo (MC) simulation of self-avoiding double-polygons.\cite{JCP}

The purpose of the present research is to study fundamental aspects of the statistical and dynamical properties of various topological polymers in solution such as the mean-square radius of gyration and the diffusion coefficient systematically through simulation.  We numerically evaluate them for two different models of topological polymers: ideal topological polymers which have no excluded volume and real topological polymers which have excluded volume, and show how they depend on the topology of the polymers. Here, the topological structures of topological polymers are expressed in terms of spatial graphs. 
Furthermore, we show that the short-range intrachain correlation is much enhanced for real topological polymers with complex graphs.

We numerically investigate fundamental properties of topological polymers in the following procedures. Firstly, we construct a weighted ensemble of ideal topological polymers by an algorithm which is based on some properties of quaternions. \cite{CDS} By the algorithm we generate $N$-step random walks connecting any given two points, \cite{CDS,UTIHD} where the computational time is linearly proportional to the step number $N$ in spite of the nontrivial constraints on the sub-chains of the graphs. We evaluate the statistical average of some physical quantities over the weighted ensemble. 
 Secondly, we construct an ensemble of conformations of real topological polymers by performing the molecular dynamics of the Kremer-Grest model and evaluate the statistical average of the  quantities over the ensemble.  
Thirdly, by comparing the results of the real topological polymers with those of the ideal ones, we numerically show that for the mean-square radius of gyration $R_G^2$ and the hydrodynamic radius $R_{H}$ the set of ratios of the quantities among different topological polymers is given by almost the same for ideal and real topological polymers if the valency of each vertex is equal to or less than three in the graphs, as far as for the polymers we have investigated. It agrees with Tezuka's observation that the SEC results are not affected by the excluded volume if each vertex of a topological polymer has less than or equal to three connecting bonds. \cite{Tezuka} 
Here we evaluate the hydrodynamic radius $R_{H}$ through Kirkwood's approximation.
 \cite{Doi-Edwards,Rubinstein,Teraoka}

It thus follows that the quaternion method for generating ideal topological polymers is practically quite useful for evaluating physical quantities. In fact, we can estimate at least approximately the values of $R_G^2$ or $R_H$ for real topological polymers by making use of the ratios among the corresponding ideal ones, if the valency of vertices is limited up to three. Moreover, we show that the ratio of the gyration radius to the hydrodynamic radius of a topological polymer, $R_G/R_H$, is characteristic to the topology of the polymer. Here, the gyration radius $R_G$ is given by the square root of the mean-square radius of gyration. We remark that the quaternion method has been generalized to a fast algorithm for generating equilateral random polygons through symplectic geometry. \cite{CS,CDSU2016,CDSUprep}

We analytically show that the ratio of the gyration radius to the hydrodynamic radius of a topological polymer, $R_G/R_H$, is characterized by the variance of the probability distribution function of the distance between two segments of the polymer. 
In particular, we argue that the ratio decreases if the variance becomes small.   Here we call the function the distance distribution function, briefly.  It is expressed with the pair distribution function, or the radial distribution function of polymer segments. The mean-square radius of gyration $R_G^2$ and the hydrodynamic radius $R_H$ correspond to the second moment and the inverse moment of the distance distribution function, respectively. 

We numerically evaluate the distance distribution function of a topological polymer both for the ideal and real chain models. For the ideal chain model, the numerical plots of the distance distribution function are approximated well by fitted curves of a simple formula. They lead to the numerical estimates of the mean-square radius of gyration and the hydrodynamic radius consistent with those evaluated directly from the chain conformations.  In the case of ideal ring polymers, we exactly derive an analytic expression of the pair distribution function, which is consistent with the fitting formula.

For real topological polymers we show that the short-distance intrachain correlation of a topological polymer is much enhanced, i.e. the correlation hole becomes large, if the graph  is complex. The exponent of the short-range power-law behavior in the distance distribution function is given by 0.7 for a linear polymer and a ring polymer, while it is given by larger values such as 0.9 and 1.15 for a $\theta$-shaped polymer and a complete bipartite graph $K_{3,3}$ polymer, respectively. We suggest that the estimate to the exponent of the short-distance   correlation such as 0.7 for linear and ring chains is consistent with the estimate of exponent $\theta_2$ of the short-range correlation in a self-avoiding walk (SAW) derived by des Cloizeaux with the renormalization group (RG) arguments, \cite{desCloizeaux} as will be shown shortly. 
Here we remark that the estimate of $\theta_2$ has been confirmed in the MC simulation of SAW. \cite{UD2013}

In order to describe the intrachain correlations of a SAW, \cite{deGennes,Schaefer} we denote by $p_0(r)$ the probability distribution function of the end-to-end distance $r$ of an $N$-step SAW. It was shown that the large-distance asymptotic behavior of $p_0(r)$ for $r > R_N$ is given by  \cite{Fisher,McKenzie} 
\begin{equation} 
p_0(r) \sim R_N^{-d} A(r/R_N) \exp\left( -(r/R_N)^{\delta} \right) \, , 
\end{equation}
where $R_N= R_0 N^{\nu}$ and $A(y)$ does not change exponentially fast for large $y$, while the short-distance behavior is given by 
\begin{equation} 
p_0(r) \sim y^g \quad (r < R_N) 
\end{equation}
with the exponent given by $g=(\gamma+1 - \alpha - d \nu)/\nu$ through asymptotic analysis \cite{McKenzie} and $g=(\gamma-1)/\nu$ through the RG arguments. \cite{desCloizeaux1974} The asymptotic properties of $p_0(r)$ are studied theoretically \cite{OO81,D85,D86} and numerically.  \cite{B81,BC91,BCRF91,Val96,TKC02,CPV02} 
In order to study the short-distance intrachain correlation between two segments of a long polymer in a good solvent, let us denote by $p_1(r)$ and $p_2(r)$ the probability distribution function between a middle point and an end point of an $N$-step SAW and that of two middle points, respectively. Then, we define critical exponents $\theta_s$ for $s=0, 1, 2$ as follows. Assuming $p_s(r) = R_N^{-d} F_s(r/R_N)$ we have 
\begin{equation} 
F_s(y) \sim y^{\theta_s} \quad (r < R_N)  \quad \mbox{for} \quad s=0, 1, 2.   
\end{equation}
The RG estimates of the exponents for $d=3$ are given by $\theta_0=0.273, \theta_1= 0.46$ and $\theta_2= 0.71$, respectively. \cite{desCloizeaux}  They are close to the MC estimates such as    
$\theta_0=0.23 \pm 0.02, \theta_1= 0.35 \pm 0.03$ and $\theta_2= 0.74 \pm 0.03$, respectively. \cite{UD2013} 
Thus, these estimates of $\theta_2$ are in agreement with the value 0.7 of the exponent 
for the short-range power-law behavior of the distance distribution functions  of real linear and real ring  polymers  within errors.  Here we remark that most of the pairs of segments in a SAW are given by those between middle points.

The contents of the paper consist of the following.  In section 2 we introduce the notation of graphs expressing topological polymers. We explain the quaternion algorithm for generating topological polymers with given graph, and then the Kremer-Grest model of molecular dynamics  and give estimates of the relaxation time. 
In section 3 we present the data of the mean-square radius of gyration and  the hydrodynamic radius both for ideal and real topological polymers. We numerically show that the ratios of the mean-square radius of gyration among such topological polymers having up to trivalent vertices are given by the same values both for ideal and real topological polymers. 
We also evaluate the ratio of the gyration radius to the hydrodynamic radius. We argue that the ratio of the root mean-square radius of gyration to the hydrodynamic radius for a topological polymer decreases if the variance of the distance distribution function decreases. 
In section 4 we evaluate the distance distribution functions both for ideal and real topological polymers. For ideal ones they are consistent with the exact result for ideal ring polymers. For the real topological polymers we numerically show that short-range intrachain correlation is much enhanced, i.e. the exponent of the short-range behavior  in the distance distribution function 
becomes large, for topological polymers with complex graphs. 
Finally, in section 5 we give concluding remarks.

%
%
\section{Topological polymers with various graphs and numerical methods}
\subsection{Definition of topological polymers} 

\begin{figure}[ht]
 \center
\includegraphics[clip,width=7.5cm]{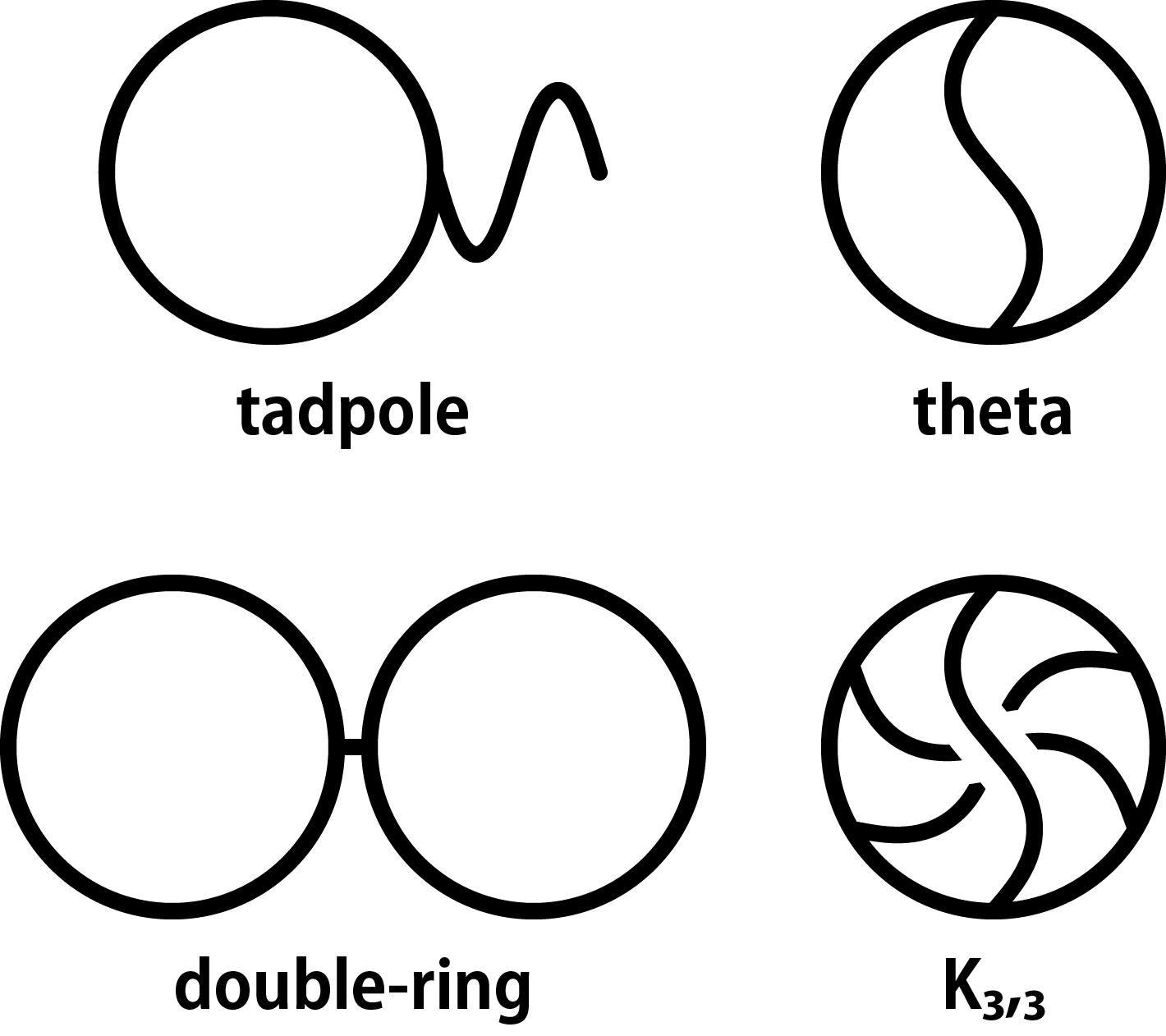}
%
 \caption{Examples of topological polymers with graphs with their names: `tadpole' for a tadpole or lasso polymer, `theta' for a $\theta$-shaped polymer, `double-ring' for a double-ring, bicyclic or 8-shaped polymer, and `$K_{3,3}$' for a complete bipartite graph $K_{3,3}$ polymer, respectively. }
\label{fig:graph}
\end{figure}

Let us call a polymer of complex topology expressed by a spatial graph $F$ 
a {\it topological polymer of graph $F$} or a {\it graph $F$ polymer}. 
For an illustration, four graphs of topological polymers are given in Fig. \ref{fig:graph}. They are tadpole, $\theta$-shaped, double-ring and complete bipartite $K_{3,3}$ graphs, respectively. 

The graph of a tadpole, which we also call a lasso, corresponds to a tadpole polymer.　It is given by a polymer of `a ring with a branch' architecture. \cite{Tezuka2001} The graph of a `theta' in Fig. \ref{fig:graph} denotes a $\theta$-shaped curve, which we also call a theta curve.  It corresponds to a $\theta$-shaped polymer, which is given by a singly-fused polymer. \cite{Tezuka2001} The graph of a `double ring' corresponds to a double-ring polymer, which we also call a di-cyclic or an 8-shaped polymer. \cite{Tezuka2001} 
Here we remark that  a complete bipartite graph $K_{3,3}$ gives one of the simplest nonplanar spatial graphs.

\subsection{Numerical method I: ideal topological polymers}

We shall explain the method of quaternions for generating 
a large number of conformations of a topological polymer expressed by a graph. \cite{CDS,UTIHD}  

\subsubsection{Hopf map of quaternions}

Let us introduce the basis $\bf i$, $\bf j$ and $\bf k$ of quaternions.  
We assume that the square of each basis  ${\bf i}, {\bf  j}$ and ${\bf k}$ is given by -1: 
\begin{equation}
{\bf i}^2 = {\bf j}^2 = {\bf k}^2 = -1 
\end{equation}
and they satisfy  the anti-commutation relations  
\begin{equation} 
{\bf i} {\bf j} = - {\bf j} {\bf i} = {\bf k} \, , \quad {\bf j} {\bf k} = - {\bf k} {\bf  j}  = {\bf i} \, , \quad {\bf k} {\bf i} = - {\bf i} {\bf k} = {\bf j} \, . 
\end{equation} 
Any quaternion $h$ is expressed in terms of the basis with real coefficients 
$w, x, y$ and $z$ as 
\begin{equation}  
h = w + x \, {\bf i} + y \, {\bf j} + z \, {\bf k} \,. \label{eq:quat}
\end{equation}
If the real part of a quaternion is given by zero,  
i.e. $w=0$ in (\ref{eq:quat}),  we call it a pure quaternion.　
We identify it with a position vector ${\vec r}=(x, y, z)$ in three dimensions.

Under the complex conjugate operation each of the basis ${\bf i}, {\bf j}$ and ${\bf k}$ changes the sign: ${\bf i}^{*} = - {\bf i}$, ${\bf j}^{*} = - {\bf j}$ and ${\bf k}^{*} = - {\bf k}$. Here we remark that the complex conjugate of the product of two quaternions is given by the product of the two complex conjugates in the reversed order: $(h_1 h_2)^{*} = h_2^{*} h_1^{*}$.

We can express any given quaternion $h$ in terms of two complex numbers $u$ and $v$ as  
\begin{equation} 
h = u + v {\bf j} \,. 
\end{equation}
We define the Hopf map by 
\begin{equation} 
{\rm Hopf}(h) = h^{*} {\bf i} h \, .  
\end{equation} 
It is instructive to show that the real part of the Hopf map vanishes \ 
\begin{eqnarray} 
h^{*} {\bf i} h 
& = & (u + v {\bf j})^{*} {\bf i} (u+ v {\bf j})  \nonumber \\ 
& = & (|u|^2 - |v|^2) {\bf i}  - 2 {\rm Im}\left( u^{*} v\right) {\bf j} 
+ 2 {\rm Re}\left(u^{*} v \right) {\bf k} \, .  
\end{eqnarray} 
Thus, the Hopf map of a quaternion gives a pure quaternion. 

For given $N$-dimensional complex vectors ${\vec u}=(u_1, u_2, \ldots, u_N)$ and 
${\vec v}=(v_1, v_2, \ldots, v_N)$ we denote by ${\vec h}$ the $N$-dimensional vector of quaternions as follows.   
\begin{equation}
{\vec h} = {\vec u} + {\vec v} \, {\bf j}. 
\end{equation}
We define the Hopf map for the $N$-dimensional vectors of quaternions 
\begin{equation} 
{\rm Hopf}({\vec h}) = {\vec h}^{\dagger} {\bf i} {\vec h} \, . \label{eq:vecHopf}
\end{equation}
For a given pair of complex vectors ${\vec u}$ and ${\vec v}$ we define bond vecters ${\vec b}_n=(x_n, y_n, z_n)$ for $n=1, 2, \ldots, N$ by 
\begin{equation}
x_n= |u_n|^2 -|v_n|^2 , \, y_n =  - 2 {\rm Im}\left( {u}_n^{*} {v_n} \right), \, 
 z_n = 2 {\rm Re}\left( {u_n}^{*} {v_n} \right)  \, . \label{eq:jumpvector}
\end{equation}
The Hopf map (\ref{eq:vecHopf}) corresponds to the sum of the bond vectors 
${\vec b}_n$'s. 

\subsubsection{Quaternion method for generating random walks with given end-to-end distance}

Suppose that a pair of Gaussian $N$-dimensional complex vectors ${\vec \alpha}$ and ${\vec \beta}$ are given, randomly. We define $N$-dimensional complex vectors ${\vec u}$ and ${\vec v}$ through the Gram-Schmidt method by 
\begin{equation}
{\vec u} = N \frac {\vec \alpha} {| {\vec \alpha} |} \, , \quad 
{\vec v} = N \frac {{\vec \beta}- ({\vec \beta} \cdot  {\vec \alpha}^{*}) 
{\vec \alpha}/{| {\vec \alpha}|^2}}  
{ |{\vec \beta} - ({\vec \beta}\cdot {\vec \alpha}^{*}) {\vec \alpha}/|{\vec \alpha}|^2| } \, . 
\label{eq:gram-schmidt}
\end{equation}
They have the same length $N$ and are orthogonal to each other with respect to the standard scalar product among complex vectors: $\langle {\vec \alpha}, {\vec \beta} \rangle = {\vec \alpha} \cdot {\vec \beta}^{*}$. Here ${\vec \alpha}^{*}$ denotes the $N$-dimensional complex vector where each entry is given by the complex conjugate of the corresponding entry of the vector ${\vec \alpha}$: $({\vec \alpha}^{*}) _n= \alpha_n^{*}$, where $\alpha_n$ are the 
$n$th component of the vector ${\vec \alpha}$. 

For a pair of complex vectors ${\vec u}$ and ${\vec v}$ constructed by (\ref{eq:gram-schmidt}) a series of the three-dimensional vectors ${\vec b}_{n}$ defined by (\ref{eq:jumpvector}) for $n=1, 2, \ldots, n$, gives a random polygon of $n$ segments. \cite{CDS} 
In order to construct such a random walk that has the end-to-end distance $R$ 
we introduce another complex vector ${\vec v}_R$ by 
\begin{equation} 
{\vec v}_R = \omega {\vec u} + \omega^{'}  {\vec v}
\end{equation} 
where weights $\omega$ and $\omega^{'}$ are given by 
\begin{equation}
 \omega = R/2N^2 \, , \quad \omega^{'} = \sqrt{1-R^2/(4 N^4)} \, .  
\end{equation}
Here we remark that the vector ${\vec v}_R$ satisfies 
\begin{equation}
{\vec u}^{\dagger} {\vec v}_R = R/2 \, . 
\end{equation}
We define another quaternion vector ${\vec h}_R$ by 
\begin{equation} 
{\vec h}_R = {\vec u} + {\vec v}_{R} 
\, {\bf j} \, .  
\end{equation}
Through the Hopf map it gives an $N$-step random walk with end-to-end distance $R$.  We have 
\begin{equation}
{\rm Hopf} ({\vec h}_{R}) = R \, {\bf k} \, . 
\end{equation} 
Thus, a series of the three-dimensional vectors ${\vec b}_{n}$ constructed by (\ref{eq:jumpvector}) with ${\vec v}$ replaced by ${\vec v}_R$ for $n=1, 2, \ldots, n$, gives a random walk of $n$ segments which connects the origin and  
the point of $z=R$ on the $z$ axis.

\subsubsection{Method for constructing random configurations of topological polymers}

We construct weighted ensembles of random configurations for the topological polymer of a graph and numerically evaluate the expectation value of a physical quantity by taking the weighted sum for the quantity. 

We first generate random configurations for the open chains and closed chains which are part of the given graph, and we attach appropriate weights to the parts of the graph. We determine the weight of the configuration for the whole graph by the product of all the weights to the parts of the graph.   

\vskip 12pt 
\par \noindent 
{\bf Theta curve graph ($\theta$-shaped graph)}
 
We generate weighted random configurations of a theta-curve graph polymer of $N=3n$ segments as follows. \cite{UTIHD} We first construct random polygons of $2n$ segments by the method of quaternions.  Secondly, we take two antipodal points A and B on the polygon such that each of the sub-chains connecting A and B has $n$ segments. Thirdly, we connect the points A and B by an $n$-step random walk with end-to-end distance equal to the distance between the points A and B by making use of the method of quaternions. 

To the whole configuration we assign the weight $w_{\rm AB}$ which is given by the $N$-step Gaussian probability density of the end-to-end vector between the points A and B. The expression of the probability density will be given by eq. (\ref{eq:ete-gauss}) in section 4.2, where the length of bond vectors is given by $b=1$.

\vskip 12 pt 
\par \noindent 
{\bf Complete bipartite graph $K_{3,3}$}

\begin{figure}[ht]
 \center
\includegraphics[clip,width=7.5cm]{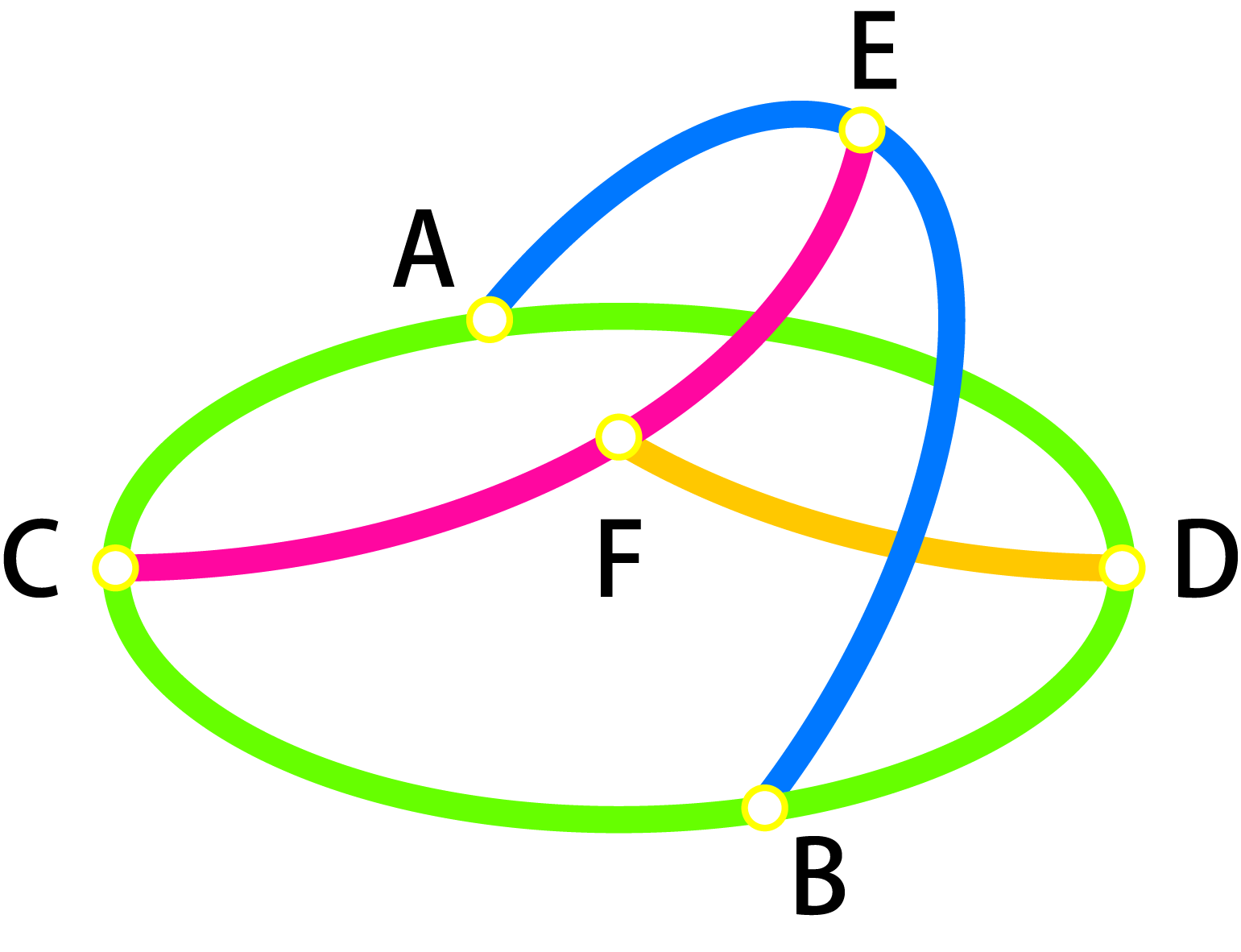}
%
 \caption{A complete bipartite graph $K_{3,3}$.  }
\label{fig:graph-complete}
\end{figure}

We generate weighted random configurations of the topological polymer of a complete bipartite graph $K_{3,3}$ with  $N=9n$ segments as follows.  
Firstly, we generate random polygons of $4n$ segments. We take antipodal points A and B on the polygon such that each sub-chain between A and B has $2n$ segments (see  Fig. \ref{fig:graph-complete}). We take a point C on one of the sub-chains between A and B such that both the sub-chain between A and C (sub-chain AC) and that of C and B (sub-chain BC) have $n$ segments, respectively.  Similarly, we take a middle point D on the other sub-chain between A and B so that sub-chains AD and BD have $n$ segments, respectively. Secondly, we generate an $2n$-step random walk such that it has the end-to-end distance equal to the distance between points A and B, and put it so that it connects A and B. We take the point E on the middle point of it so that sub-chains AE and BE have $n$ segments, respectively. Thirdly, we generate a $2n$-step random walk connecting C and E. We take the middle point $F$ on the walk so that sub-chains CF and EF have $n$ segments, respectively. Finally, we generate an $n$-step random walk DF which connects points D and F.   
 
We associate the weight $w_{\rm AB}$, $w_{\rm CE}$ and $w_{\rm FD}$ for sub-chains AB, CE and FD, respectively. We define the weight for the whole configuration by multiplying them as  $w_{\rm total}= w_{\rm AB} w_{\rm CE} w_{\rm FD}$.  

\subsection{Numerical Method II: Molecular dynamics simulation}

\subsubsection{The Kremer-Grest model}

A polymer chain in the Kremer-Grest model has both the repulsive Lennard-Jones (LJ) potentials and the finitely extensible nonlinear elongation (FENE) potentials to prevent the bonds from crossing each other. 
The LJ potential is given by
\begin{equation}
\label{eq_GeneralLJ}
E_{LJ}\left(r_{ij}\right)=4\epsilon\left(\left(\frac{\sigma}{r_{ij}}\right)^{12}-\left(\frac{\sigma}{r_{ij}}\right)^6\right)
\end{equation}
where $r_{ij}$ is the distance between the $i$th and $j$th atoms and we set the LJ parameters $\epsilon$ and $\sigma$ as $\epsilon=1.0$ and $\sigma=1.0$. The term of $r_{ij}^{-12}$ corresponds to the short-range repulsion, and that of $r_{ij}^{-6}$  the long-range attractive interaction. The minimum of the LJ potential is given by $-1$ at $r_{ij}=2^{1/6}\approx1.122$. We introduce cutoff in order to produce a repulsive Lennard-Jones potential as follows.
\begin{equation}
\label{eq_RepulsiveLJ}
E^{'}_{LJ}\left(r_{ij}\right)=\left\{
\begin{array}{ll}
4\left(\left(\frac{1}{r_{ij}}\right)^{12}-\left(\frac{1}{r_{ij}}\right)^6+1/4\right) & \left(r_{ij}<2^{1/6}\right) \\
0 & \left(r_{ij}>2^{1/6}\right) \\
\end{array}
\right.
\end{equation}
where the constant term $1/4$ is added to eliminate the discontinuous jump at $r_{ij}=2^{1/6}$.

The FENE potential between a pair of bonded atoms
\begin{equation}
\label{eq_fene}
E_{F}\left(r_{ij}\right)=-0.5K_0R_0^2\log \left[1-\left(\frac{r_{ij}}{R_0}\right)^2\right]
\end{equation}
is employed to provide a finitely extensible and nonlinear elastic potential, $R_0$ is the maximum extent of the bond. Here we choose $K_0=30.0$ and $R_0=1.5$.

\subsubsection{Relaxation time for the evolution through LAMMPS}

We generate an ensemble of conformations of a topological polymer of graph $F$ by LAMMPS: The initial conformation is given by putting the atoms on the lattice points along the polygonal lines of the given graph $F$ in a cubic lattice. Then by LAMMPS we integrate Newton's equation of motion for the atoms under the repulsive Lennard-Jones and FENE potentials.

The topological type of the conformation of a topological polymer does not change during time evolution. The bonds can hardly cross each other, since the atoms are surrounded by strong barriers which increase with respect to the inverse of $r_{ij}^{12}$ while they are connected with nonlinear elastic springs of finite length.

Let us denote by ${\vec r}_j$ the position vectors of the $j$th segments of a polymer for $j=1, 2, \ldots, N$.  We define the correlation between the conformation at time $t$ and that of the initial time $t=0$ by 
\begin{equation}
\label{eq_correlation}
\frac{\sum_{i=1}^N \left(\vec{r}_i(t)-\vec{R}_c(t)\right) \, \cdot \, \left(\vec{r}_i(0)-\vec{R}_c(0)\right)}{\sqrt{\sum_{i=1}^N \left(\vec{r}_i(t)-\vec{R}_c(t)\right)^2}\sqrt{\sum_{i=1}^N \left(\vec{r}_i(0)-\vec{R}_c(0)\right)^2}} \, , 
\end{equation}
where the center of mass of the polymer, ${\vec R}_c$, is given by  
\begin{equation}
{\vec R}_c = {\frac 1 N} \sum_{j=1}^{N} {\vec r}_j \, .  
\end{equation} 
We define the relaxation time $\tau$ of the conformational correlation by the number of steps at which the conformational correlation decreases to $1/e$.

We take independent conformations at every $6\tau$ steps. 
For example, we have evaluated $\tau=6180$ for a linear polymer of $N=500$ segments,  
$\tau= 3800$ for a ring polymer of $N=500$ segments, and $\tau=2430$ for a $K_{3,3}$ graph polymer of $N=492$ segments.

\subsection{Numerical Method III: Distribution function of the distance between two segments}

Let us denote the distance distribution function by $F(r)$ in terms of  distance $r$. 
Here we recall that it is the probability distribution function of the distance between any given pair of segments of a polymer. The probability that the distance between a given pair of segments in the polymer is larger than $r$ and less than $r + dr$ 
is given by $F(r) dr$.  Suppose that there are $N$ segments in a region of volume $V$. 
We denote the average global density by $\rho=N/V$.  
In terms of the pair distribution function $\rho g(r)$,  
it is expressed as $N F(r) dr = 4 \pi r^2 g(r) dr$.   

We numerically evaluate the distance distribution function $F(r)$ as follows. Firstly, we generate an ensemble of random conformations of a polymer with graph $F$. Secondly, for each conformation we choose a pair of segments randomly, and calculate the distance between them. We repeat the procedure several times. Thirdly, we make the histogram of the estimates with the distance between two segments for all conformations in the ensemble.

%
%
\section{Mean-square radius of gyration and the hydrodynamic radius}
\subsection{Mean-square radius of gyration for topological polymers}

 We define the mean-square radius of gyration for a topological polymer of graph $F$ consisting of $N$ segments by    
\begin{equation} 
\langle R_G^2 \rangle_F = {\frac 1 N} \sum_{j=1}^{N} \langle ({\vec r}_j - {\vec R}_c)^2 \rangle_F \, .  
\end{equation} 
Here the symbol  $\langle A \rangle_F$  denotes the ensemble average of $A$  
over all  possible configurations of the topological polymer with graph $F$.    
We denote by $R_G(F)$ the square root of the mean-square radius of gyration 
$\langle R_G^2 \rangle_F$: 
\begin{equation} 
R_G(F)= \sqrt{ \langle R_G^2 \rangle_F} \, .
\end{equation} 
 We also call it the gyration radius of the polymer.

\begin{figure}[ht]
 \center
\includegraphics[clip,width=8.0cm]{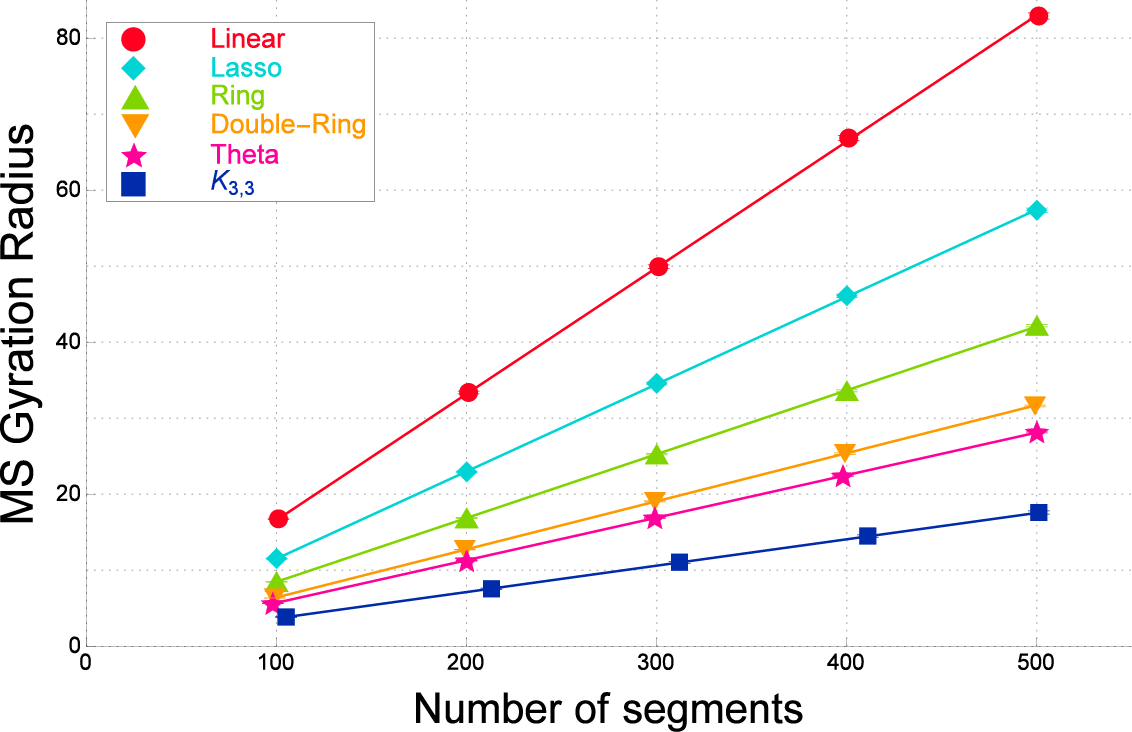}
%
 \caption{Mean-square radius of gyration $\langle R_G^2 \rangle_F$ versus the number of segments $N$ for ideal topological polymers with graph $F$  evaluated by the quaternion method for  six graphs $F$ such as linear, tadpole (lasso), ring, double-ring, $\theta$-shaped, and  complete bipartite graph $K_{3,3}$ polymers, depicted by filled circles, filled diamonds, filled upper triangles, filled lower triangles, filled stars, and filled squares, respectively.  Each data point corresponds to the average over $10^4$ samples. }
\label{fig:QRg}
\end{figure}

In Fig. \ref{fig:QRg} we plot against the number of segments $N$ the numerical estimates of the mean-square radius of gyration $\langle R_G^2 \rangle_F$ for ideal topological polymers of graph $F$ for six different graphs. They are evaluated by the quaternion method, and are given in decreasing order for a given number of segments $N$ as follows: those of linear polymers, tadpole (lasso) polymers, ring polymers, double-ring polymers, $\theta$-shaped polymers, and polymers with a complete bipartite graph $K_{3,3}$. Here we remark that the markers in Figures correspond to the same graphs throughout the paper.

We observe in Fig. \ref{fig:QRg} that the estimate of the gyration radius for a double-ring polymer of $N$ segments, depicted by filled lower triangles, is close to that of a $\theta$-shaped polymer of $N$ segments depicted with filled stars. The former is only slightly larger than the latter. For various other graphs the estimates of the mean-square radius of gyration are distinct among the different graphs with same given number of segments $N$. 

The estimates of the mean-square radius of gyration for the ideal topological polymers linearly depend on the number of segments $N$, as shown in Fig. \ref{fig:QRg}. They are fitted by the formula 
\begin{equation} 
\langle R_G^2 \rangle_F = a_0 + a_1 N \, . \label{eq:fit-ideal}
\end{equation}
The best estimates of parameters $a_0$ and $a_1$ 
in (\ref{eq:fit-ideal}) are listed in Table \ref{tab:ideal-Rg} together with $\chi^2$ values per degree of freedom (DF). 
The $\chi^2$ values per DF for the topological polymers with the different graphs are at most 1.1 and are small.

\begin{table}[htbp] 
\begin{tabular}{c|ccc}
graph $F$ &  $a_0$ & $a_1$ & $\chi^2$/DF \\ 
\hline  
linear & $-0.004 \pm 0.081$ & $0.16605 \pm 0.00043$ & $0.37$  \\ 
tadpole 
(lasso)
 & $0.039 \pm 0.059$ & $0.11483 \pm 0.00032$ & $0.43$ \\ 
ring & $0.084 \pm 0.025$ & $0.08397 \pm 0.00014$ & $0.29$  \\ 
double-ring & $0.100 \pm 0.020$ & $0.06327 \pm 0.00011$ & $0.33$  \\ 
$\theta$-shaped & $0.120 \pm 0.037$ & $0.05600 \pm 0.00021$ & $1.13$  \\ 
complete $K_{3,3}$ & $0.191 \pm 0.022$ & $0.03470 \pm 0.00011$ & $0.09$  \\ 
\hline 
\end{tabular} 
\caption{Best estimates of the parameters in eq.  (\ref{eq:fit-ideal}) fitted to the estimates of  the mean-square radius of gyration $\langle R_G^2 \rangle_F$ for ideal topological polymers of graph $F$ evaluated by the quaternion method and the $\chi^2$ values per degree of freedom (DF). }
\label{tab:ideal-Rg}
\end{table}

\begin{figure}[ht]
 \center
\includegraphics[clip,width=8.0cm]{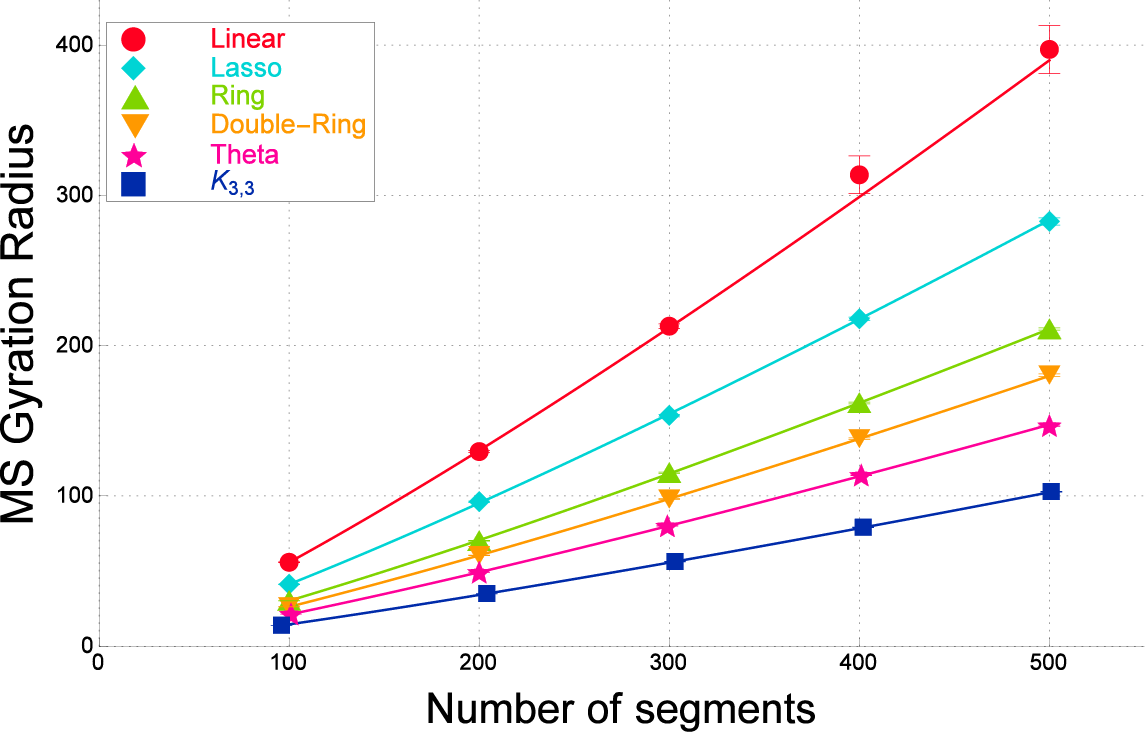}
%
 \caption{Mean-square radius of gyration $\langle R_G^2 \rangle_F$ versus the number of segments $N$ for real topological polymers with graphs $F$ evaluated by the 
molecular dynamics of the Kremer-Grest model. 
Each data point corresponds to the average over $5 \times 10^3$ samples. }
\label{fig:LAMMPS_RG}
\end{figure}

For real topological polymers, which have excluded volume, the numerical values of the mean-square radius of gyration are proportional to some  power of the number of segments $N$ with scaling exponent $\nu=0.59$, if the number of segments $N$ is large enough. We derive good theoretical curves fitting to the data by applying the following formula:   
\begin{equation} 
\langle R_G^2 \rangle_F = a_0 + a_1 \, N^{2 \nu} \, .  \label{eq:fit-real}
\end{equation}
Here we fix the value of scaling exponent $\nu$ by $\nu=0.59$ when we apply formula (\ref{eq:fit-real}) to the data in order to derive the best estimates of parameters $a_0$ and $a_1$. 

We observe in Fig. \ref{fig:LAMMPS_RG} that the estimate of the gyration radius 
for a double-ring polymer of $N$ segments is  clearly larger  than  that  of a $\theta$-shaped polymer of $N$ segments, for each $N$. The former and the latter are distinct for each segment number $N$. It is in agreement with Tezuka's observation in experiments that if a polymer has such a vertex with four connecting bonds, the average size of the polymer is much enhanced due to the excluded volume effect. \cite{Tezuka} Here we remark that a double-ring polymer indeed has a vertex where four bonds are connected. 
We recall that in the case of ideal topological polymers the size of double-ring polymers is close to that of $\theta$-shaped polymers with the same number of segments $N$.  

It has been shown in the MC simulation \cite{JCP} that in the case of nonzero finite excluded volume the mean-square radius of gyration for double-ring polymers of $N$ segments is much closer to that of ring polymers of $N$ segments, while in the case of no excluded volume the mean-square radius of gyration for double-ring polymers of $N$ segments is distinctly smaller than that of ring polymers of $N$ segments. 

\begin{table}[htbp] 
\begin{tabular}{c|ccc} 
graph $F$ & $a_0$ & $a_1$ & $\chi^2$/DF \\ 
\hline 
linear & $-3.02 \pm 0.62$ & $0.2568 \pm 0.0019$ & $0.99$  \\ 
tadpole (lasso) & $-1.42 \pm 0.43$ & $0.1863 \pm 0.0011$ & $1.86$  \\ 
ring & $-1.57 \pm 0.19$ & $0.13903 \pm 0.00047$ & $1.80$  \\ 
double-ring & $-0.834 \pm 0.064$ & $0.11813 \pm 0.00016$ & $0.22$  \\ 
$\theta$-shaped & $-1.13 \pm 0.10$ & $0.09711 \pm 0.00026$ & $2.22$  \\ 
complete $K_{3,3}$ & $-0.883 \pm 0.017$ & $0.067521 \pm 0.000051$ & $0.28$  \\ 
\hline 
\end{tabular} 
\caption{Best estimates of the parameters in eq. (\ref{eq:fit-real}) fitted to the data of the mean-square radius of gyration for topological polymers with excluded volume (the Kremer-Grest model) and the $\chi^2$ values/DF.}
\label{tab:LAMMPS_RG}
\end{table}

We now argue that the ratios among the estimates of the mean-square radius of gyration for ideal topological polymers with graphs $F$ are approximately similar to those of real topological polymers with the same such graphs $F$ consisting of at most trivalent vertices, as far as we have investigated. 

Let us first consider the estimates of coefficient $a_1(F)$ in formula (\ref{eq:fit-ideal}) for ideal topological polymers with graphs $F$.  If $N$ is large enough, the coefficients $a_1$ determine the ratios among the gyration radii $R_G(F)$ for some graphs $F$.  We thus calculate the ratio $a_1(F)/a_1({\rm ring})$ for ideal topological polymers of graph $F$ in order to study the ratios among the gyration radii of topological polymers with different graphs. 
They are listed in table \ref{tab:ratio}. 
We then consider the estimates of coefficient $a_1(F)$ in formula (\ref{eq:fit-real}) for real topological polymers with graphs $F$. We again calculate the ratios $a_1(F)/a_1({\rm ring})$  for those of real topological polymers. They are listed in the third column of Table \ref{tab:ratio}.

\begin{table}[htbp] 
\begin{tabular}{c|cc} 
graph $F$ & ratio of $a_1$ (ideal) & ratio of $a_1$ (real)   \\ 
\hline 
linear & $1.978 \pm 0.015$ & $1.847 \pm 0.038$  \\ 
tadpole (lasso) & $1.368 \pm 0.009$ & $1.340 \pm 0.018$   \\ 
ring & $1.000 \pm 0.003$ & $1.000 \pm 0.007$   \\ 
double-ring & $0.753 \pm 0.002$ & $0.850 \pm 0.002$   \\ 
$\theta$-shaped & $0.667 \pm 0.004$ & $0.699 \pm 0.003$  \\ 
complete $K_{3,3}$ & $0.413 \pm 0.002$ & $0.4857 \pm 0.0005$   \\ 
\hline 
\end{tabular} 
\caption{Ratio of coefficients, $a_1(F)/a_1({\rm ring})$, in eqs. (\ref{eq:fit-ideal}) and (\ref{eq:fit-real}) for ideal and real topological polymers with graphs $F$, respectively. }
\label{tab:ratio}
\end{table} 

For ideal polymers the ratio $a_1({\rm linear})/a_1({\rm ring})$ is given by 1.98, which is almost equal to 2 with respect to errors, while for real polymers it is given by 1.84. It is clearly smaller than 2 with respect to errors.  Thus, the ratio $a_1({\rm linear})/a_1({\rm ring})$  is smaller than 2 for real polymers.

 We suggest that for the real linear and ring polymers the ratio $a_1({\rm linear})/a_1({\rm ring})$  is smaller than 2 due to the excluded volume effect. In fact, the mean-square radius of gyration is enhanced through the effective repulsions among segments under the excluded volume. We therefore conclude that the excluded volume is more important in ring polymers than in linear polymers. We suggest that the swelling effect of ring polymers is more significant than in the linear polymers. 

For other graphs $F$ such as tadpole and $\theta$-shaped polymers we observe in Table \ref{tab:ratio} that ratio $a_1(F)/a_1({\rm ring})$ is given by almost the same value both for ideal and real topological polymers. 
Here we remark that the two graphs have at most three connecting bonds at all vertices. However, in the case of double-ring polymers the ratio is different for ideal and real polymers.  The behavior of ratio $a_1(F)/a_1({\rm ring})$ is in agreement with the previous observation 
in Figs. \ref{fig:QRg} and \ref{fig:LAMMPS_RG} that the ratio of the mean-square radius of gyration for real double-ring polymers to that of real $\theta$-shaped polymers is clearly larger than that of the corresponding ideal topological polymers.  
For the complete bipartite graph $K_{3,3}$ polymer  the ratio  $a_1(K_{3,3})/a_1({\rm ring})$ for real polymers is larger than that of ideal polymers. It is possible that the number of segments $N=500$ is not large enough to investigate the asymptotic behavior of a polymer with such a complex graph since each sub-chain has only about 55 segments for $N=500$. It is necessary to perform simulation for much larger $N$ in order to study the asymptotic behavior.

The excluded volume plays a significant role in a double-ring polymer, for which one vertex has four valency. \cite{JCP} In fact, the ratio $a_1({\rm double} \, \, {\rm ring})/a_1({\rm ring})$ for a real double-ring polymer is distinctly larger than that of an ideal double-ring polymer. In Table \ref{tab:ratio} the former is given by 0.85 while for the latter given by 0.75. The former is distinctly larger than the latter with respect to errors.

\subsection{Diffusion coefficient via the hydrodynamic radius for a topological polymer}

We evaluate the diffusion coefficients of topological polymers in solution by    applying Kirkwood's approximation. \cite{Doi-Edwards,Rubinstein,Teraoka} The diffusion coefficient, $D(F)$, of a topological polymer with graph $F$ consisting of $N$ segments  in a good solvent with viscosity $\eta$ at a temperature $T$ is given by 
\begin{equation} 
D(F) = \frac {k_B T} {6 \pi \eta } {\frac 1 {N^2}} \sum_{n,m=1; n \ne m}^{N} 
\langle \frac 1 {|\vec{ r}_m - \vec{r}_n|} \rangle_F  \, .  
\end{equation}
Here we take the average of the inverse distance between two segments over all pairs of segments of a given topological polymer. Thus, it is useful to introduce the hydrodynamic radius $R_H(F)$ of a topological polymer with graph $F$ 
of $N$ segments by 
\begin{equation}
{\frac 1 {R_H(F)}} =  {\frac 1 {N^2}} \sum_{n,m=1; n \neq m}^{N} 
\langle \frac 1 {|\vec{ r}_m - \vec{r}_n|} \rangle_F   \, .  \label{eq:RH}
\end{equation} 

\begin{figure}[ht]
\center
\includegraphics[clip,width=8.0cm]{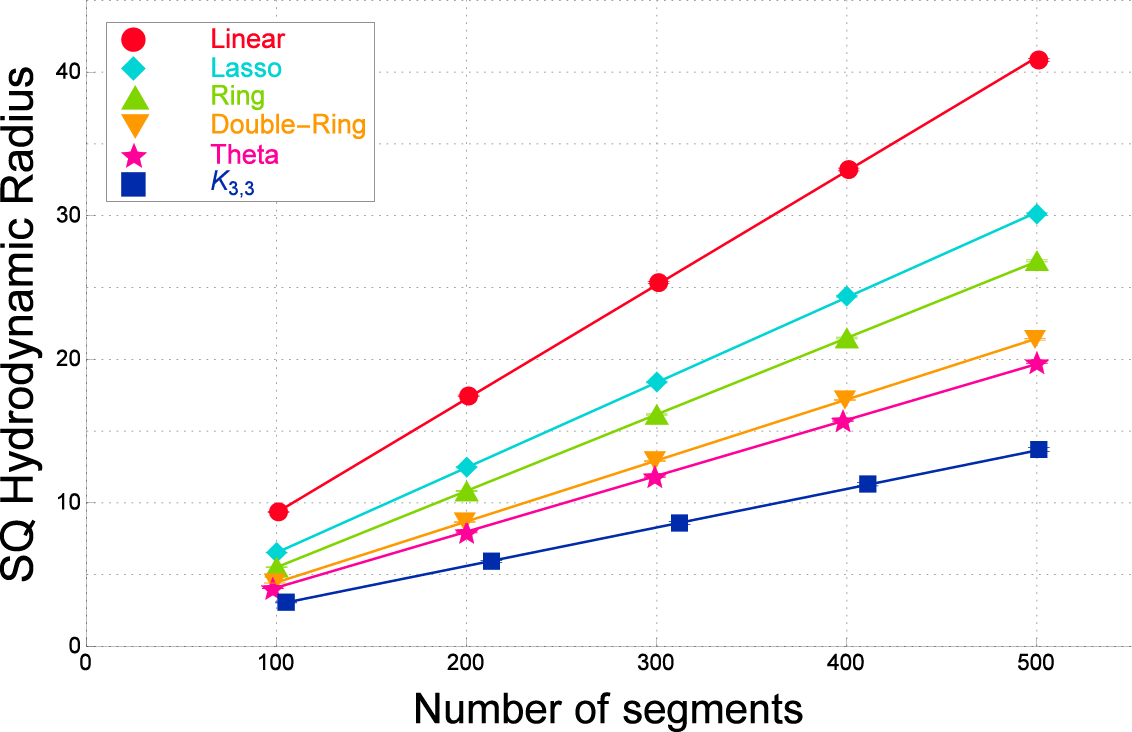}
%
%
\caption{Hydrodynamic radius squared $\left( R_H(F) \right)^2$ versus the number of segments $N$ for ideal topological polymers of graphs $F$ evaluated  by the quaternion method.}
\label{fig:QRH}
\end{figure}

In Fig. \ref{fig:QRH} we plot against the number of segments $N$ the square of the hydrodynamic radius, $\left( R_H(F)\right)^2 $, for ideal topological polymers with graphs $F$ evaluated by the quaternion method. 
%
The curves fitted to the data points are given by applying the following formula  
\begin{equation} 
\left( R_H(F) \right)^2 = a_0 + a_1 N \, .  \label{eq:fit-QRH}
\end{equation}
The best estimates of the parameters $a_0$ and $a_1$ are listed in Table \ref{tab:QRH}. 
\begin{table}[htbp]
\begin{tabular}{c|ccc}
graph $F$ & $a_0$ &  $a_1$  & $\chi^2$/DF \\  \hline
linear & $1.404 \pm 0.085$ & $0.07927 \pm 0.00042$ & $3.96$  \\
tadpole (lasso) & $0.611 \pm 0.033$ & $0.05926 \pm 0.00017$ & $1.34$  \\
ring & $0.193 \pm 0.013$ & $0.053242 \pm 0.000066$ & $0.29$  \\
double-ring & $0.2141 \pm 0.0058$ & $0.042469 \pm 0.000030$ & $0.1$  \\
$\theta$-shaped & $0.211 \pm 0.030$ & $0.03893 \pm 0.00016$ & $1.88$  \\
complete $K_{3,3}$ & $0.246 \pm 0.021$ & $0.02682 \pm 0.00010$ & $0.15$  \\
\hline 
\end{tabular}
\caption{
Best estimates of the parameters in eq. (\ref{eq:fit-QRH}) for curves fitted to the hydrodynamic radius $R_H$ of ideal topological polymers versus the number of segments $N$ for various graphs and the $\chi^2$ values/DF. }
\label{tab:QRH}
\end{table}

We observe in Fig. \ref{fig:QRH} that the $N$-dependence of the hydrodynamic radius is well approximated by the fitted curves of formula (\ref{eq:fit-QRH}).
However, the curves fitted to the data are not very good since the $\chi^2$/DF are not very small. In fact, if we apply a simple power of $N$ to the data such as $N^{\nu}$, the estimates of scaling exponent $\nu$ are less than 0.5, as shown in Appendix A.  
In experiments the data points of the hydrodynamic radius versus the number of segments for linear polymers in a $\theta$ solvent are fitted by a curve with exponent $\nu=0.484$. \cite{Teraoka,Suda}

We also observe in Fig. \ref{fig:QRH} that the hydrodynamic radius $R_H(F)$ of an ideal double-ring polymer of $N$ segments is close to but slightly larger than 
that of an ideal $\theta$-shaped polymer of $N$ segments. It is similar to the case of the mean-square radius of gyration.

\begin{figure}[ht]
\center
\includegraphics[clip,width=8.0cm]{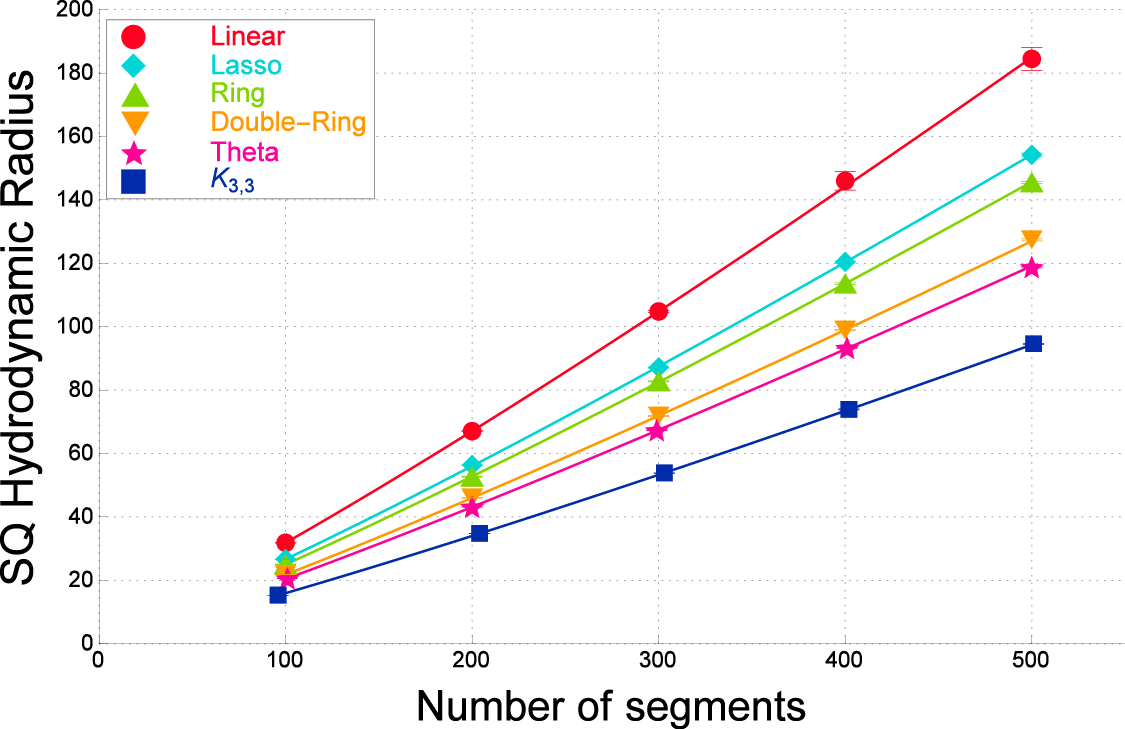}
%
 \caption{Hydrodynamic radius squared $\left( R_H(F) \right)^2$ 
versus the number of segments $N$ for real topological polymers of graphs $F$ 
 (the Kremer-Grest model). }
\label{fig:LRH}
\end{figure}

In Fig. \ref{fig:LRH} we plot against the number of segments $N$ the hydrodynamic radius for real topological polymers evaluated by performing the molecular dynamical simulation of the Kremer-Grest model through LAMMPS. Here we recall that each molecule has excluded volume in the model.  
The fitted curves are given by applying formula 
(\ref{eq:fit-real})  to the data points where we put $\nu=0.565$  
\begin{equation} 
\left( R_H(F)  \right)^2 = a_0 + a_1 N^{1.13} \, .  \label{eq:fit-LRH}
\end{equation}
The best estimates of the parameters $a_0$ and $a_1$ are listed in  
Table \ref{tab:LRH}.  Here, the value of the exponent 
$\nu= 0.565$ is slightly smaller than the scaling exponent of SAW:  
$\nu_{\rm SAW} = 0.588$. 
However, it is also the case in polymer experiments such as $\nu=0.567$. \cite{Teraoka,Bishop}

We observe in Fig. \ref{fig:LRH} that the hydrodynamic radius $R_H$ of real double-ring polymers is approximately close to but distinctly larger than that of ideal $\theta$-shaped polymers. The enhancement due to the excluded volume is not as significant as in the case of the mean-square radius of gyration. However, the ratio of estimates of $a_1$ for double-ring and $\theta$-shaped polymers is clearly larger than 1 as we have $a_1({\rm double} \, \, {\rm ring})/a_1({\rm theta}) = 1.065$.

\begin{table}[htbp]
\begin{tabular}{c|ccc}
graph $F$ & $a_0$ &  $a_1$  & $\chi^2$/DF \\  \hline
linear & $2.204 \pm 0.059$ & $0.16292 \pm 0.00022$ & $0.14$  \\
tadpole (lasso) & $1.98 \pm 0.15$ & $0.13578 \pm 0.00045$ & $2.73$  \\
ring & $1.653 \pm 0.075$ & $0.12843 \pm 0.00022$ & $1.01$  \\
double-ring & $1.438 \pm 0.054$ & $0.11193 \pm 0.00017$ & $0.81$  \\

$\theta$-shaped & $1.278 \pm 0.063$ & $0.10511 \pm 0.00019$ & $1.66$  \\  
complete $K_{3,3}$ & $0.796 \pm 0.016$ & $0.083379 \pm 0.000058$ & $0.35$  \\
\hline 
\end{tabular}
\caption{Best estimates of the parameters in eq.(\ref{eq:fit-LRH}) for the hydrodynamic radius $R_H(F)$  of real topological polymers with graph $F$ of $N$ segments for various graphs $F$. } 
\label{tab:LRH} 
\end{table}

\subsection{Ratio of the gyration to hydrodynamic radii}

Let us now consider the ratio of the gyration radius $R_G$ to the hydrodynamic radius $R_H$ 
for topological polymers with graphs $F$. We suggest that the ratio is universal in the sense of the renormalization group arguments, i.e., we expect that it is independent of some details of the polymer model,  so that we can compare the estimate of the ratio with other theoretical approaches and experiments.  Here we take the analogy of the amplitude ratio of susceptibilities and that of specific heats in critical phenomena. \cite{deGennes} 

\begin{figure}[ht]
 \center
\includegraphics[clip,width=8.5cm]{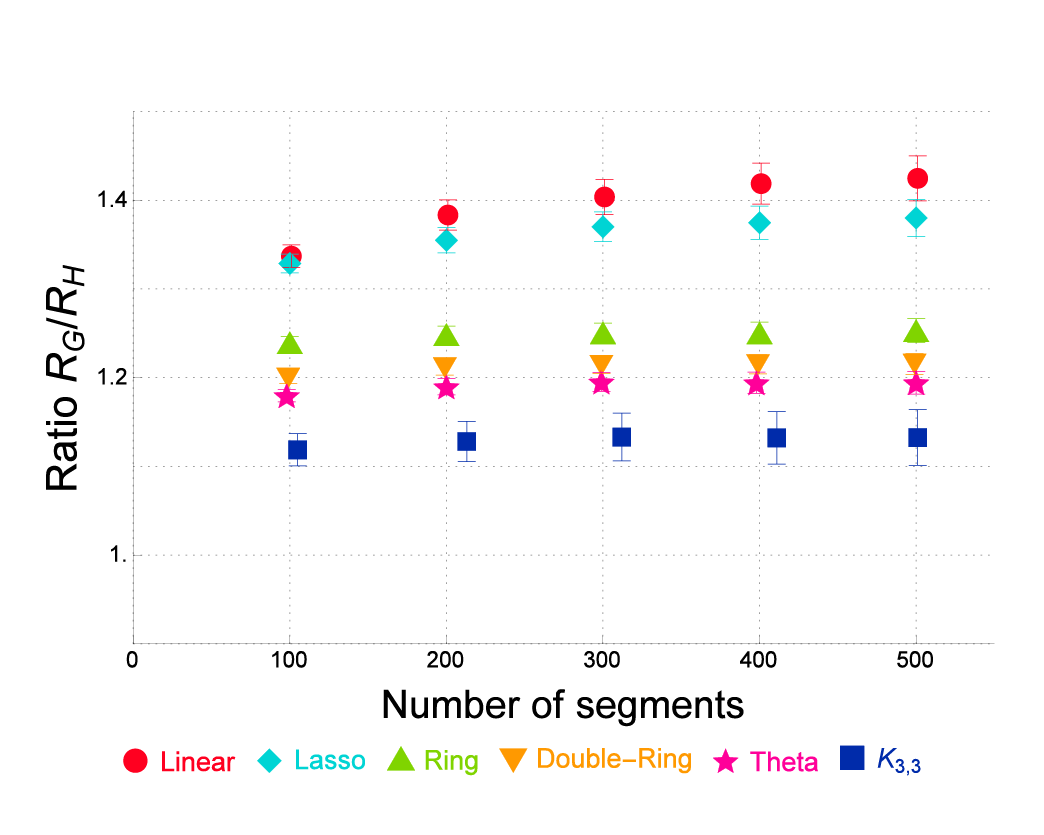}
%
%
 \caption{Ratio of the gyration radius to the hydrodynamic radius, $R_G(F)/R_H(F)$, 
versus the number of segments $N$   
for ideal topological polymers with graphs $F$ evaluated  by the quaternion method}
\label{fig:ratio-Q}
\end{figure}

\begin{figure}[ht]
 \center
\includegraphics[clip,width=8.5cm]{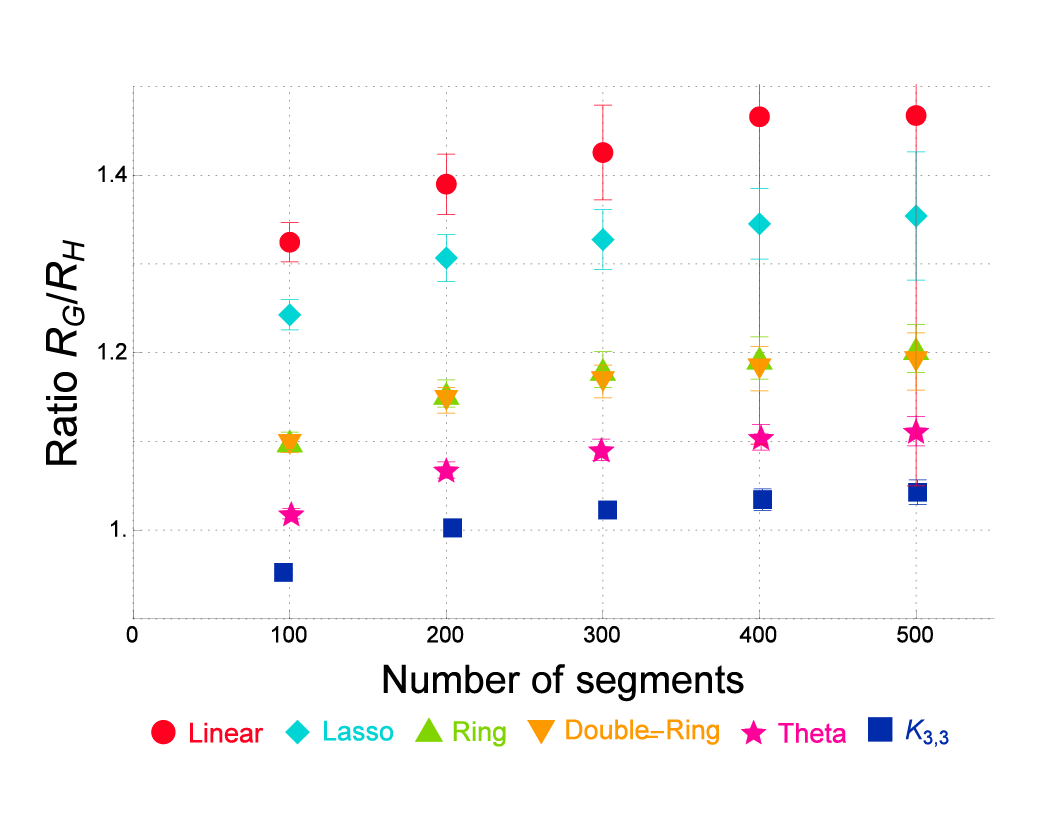}
%
\caption{Ratio of the gyration radius to the hydrodynamic radius 
$R_G(F)/R_H(F)$ versus the number of segments $N$ 
for real topological polymers (the Kremer-Grest model). }
\label{fig:ratio-L}
\end{figure}

We numerically evaluated the ratio $R_G/R_H$ for several topological polymers with various different graphs $F$.
For ideal topological polymers of graphs $F$  the numerical estimates of the ratio $R_G/R_H$ 
are plotted against the number of segments $N$ in Fig. \ref{fig:ratio-Q}.  
For real topological polymers with graph $F$ (the Kremer-Grest model)  
the estimates of the ratio of the gyration radius
to the hydrodynamic radius,  $R_G(F)/R_H(F)$,  
are plotted against the number of segments $N$ in Fig. \ref{fig:ratio-L}.

By comparing Figs. \ref{fig:ratio-Q} and \ref{fig:ratio-L} we observe that the estimates of the ratio of the gyration radius to the hydrodynamic radius are given in the same order for both the ideal and the real topological polymers of graphs $F$. In decreasing order the ratios are given by that of a linear polymer, tadpole (lasso) polymer, ring polymer, double-ring polymer, $\theta$-shaped polymer, and the polymer of a complete bipartite graph $K_{3,3}$, for ideal topological polymers as shown in Fig. \ref{fig:ratio-Q}, and it is also the case for 
real topological polymers as shown in Fig. \ref{fig:ratio-L},   

In each of Figs. \ref{fig:ratio-Q} and \ref{fig:ratio-L}  the ratio $R_G(F)/R_H(F)$ increases with respect to the 
number of segments $N$ gradually and monotonically. 
As the number of segments $N$ becomes very large such as $N=500$, 
the ratios approach constant values for all graphs $F$ as shown in  Figs. \ref{fig:ratio-Q} and \ref{fig:ratio-L}.   
The ratios among topological polymers with the graphs $F$ are given by almost the same values both for the ideal topological polymers and the real topological polymers. 

The ratios for topological polymers with various graphs $F$ at $N=500$ are listed in Table \ref{tab:ratioN500}. In each case of graphs $F$   the ratio for the ideal topological polymer 
and that of the real topological polymer is given by almost the same values with respect to errors. We suggest that the estimates of the ratio $R_G(F)/R_H(F)$ should be universal since they are given by almost the same value for different models.

\begin{table}[htbp] 
\begin{tabular}{c|ccc} 
graph $F$ & $N$ & $R_G(F)/R_H(F)$ (ideal) & $R_G(F)/R_H(F)$ (real)  \\
\hline 
linear & 500 & $1.425 \pm 0.025$ & $1.47 \pm 0.42$ \\ 
tadpole (lasso) & 500 & $1.380 \pm 0.021$ & $1.354 \pm 0.072$  \\ 
ring & 500 & $1.253 \pm 0.013$ & $1.205 \pm 0.027$   \\ 
double-ring &  500 & $1.215 \pm 0.011$ & $1.190 \pm 0.032$  \\ 
$\theta$-shaped & 500 &  $1.194 \pm 0.013$ & $1.112 \pm 0.017$  \\ 
complete $K_{3,3}$ & 501 & $1.132 \pm 0.031$ & $1.043 \pm 0.014$  \\ 
\hline 
\end{tabular} 
\caption{Ratio of the gyration radius to the hydrodynamic radius $R_G/R_H$ 
versus the number of segments $N$ for ideal and real topological polymers with graph $F$.  }
\label{tab:ratioN500}
\end{table}

\subsection{Ratio $R_G/R_H$ in terms of the mean-square deviations of the distance between two segments}

Let us denote by ${\bar r}$ and $\sigma^2$ the mean value  and the variance of the distance $r$ between randomly chosen pairs of segments of a polymer, respectively. We now argue that the ratio of the gyration radius to the hydrodynamic radius, $R_G/R_H$, decreases if the standard deviation $\sigma$ becomes small compared with the mean distance ${\bar r}$.

We first recall that the mean-square radius of gyration of a polymer is given by the average of the square distance between two segments over all the pairs    
\begin{equation} 
R_G^2 = \frac 1 {2N^2} \sum_{j, k=1}^{N} \langle \left( {\vec r}_j - {\vec r}_k \right)^2 \rangle \, . 
\end{equation} 
Therefore, the mean-square radius of gyration of a polymer is expressed in terms of the mean-square distance between two segments of the polymer, $\langle r^2 \rangle$, as follows.  
\begin{equation}
R_G^2 = \langle r^2 \rangle /2 \, . 
\end{equation}
Through Kirkwood's approximation the hydrodynamic radius is given by  
the average of the inverse distance between two segments over all the pairs of segments   
\begin{equation}
R_H^{-1} = \langle 1/r \rangle \, . 
\end{equation}
Thus, the gyration radius is given by the root of the mean-square of the distance between two segments, while the hydrodynamic radius by its harmonic mean.

Let us express distance $r$ in terms of the mean distance ${\bar r}$ 
and the deviation $\Delta r$ as 
$r= {\bar r} + \Delta r$. 
We define the variance $\sigma^2$ of distance $r$ by 
 \begin{equation}
\sigma^2 = \langle (\Delta r)^2 \rangle \, . 
\end{equation} 
The  mean-square of distance $r$ is given by  
\begin{eqnarray} 
\langle r^2 \rangle & = & \langle ({\bar r} + \Delta r)^2 \rangle \nonumber \\ 
& = & {\bar r}^2 + \sigma^2 \, . 
\end{eqnarray} 
If the mean distance $\bar r$ is much larger than the standard deviation $\sigma$ of the distance $r$, we can expand the inverse distance as a series of $\Delta r$: 
$1/r = 1/{\bar r} - \Delta r/{\bar r}^2 + \cdots  $. 
We have the following approximation 
\begin{equation}
\langle 1/r \rangle = 1/ {\bar r}  + \sigma^2/{\bar r}^3 \, . 
\end{equation}
The ratio of the gyration radius to the hydrodynamic radius is therefore 
approximately given by 
\begin{equation} 
R_G/R_H = {\frac 1 {\sqrt{2}}} \,  \left( 1 + {\frac {\sigma^2} {{\bar r}^2}} \right)^{3/2} \, .  
\end{equation}

%
%
\section{Pair distribution function of a topological polymer}
\subsection{Pair distribution function}

Let us define the local segment density $\rho({\vec r})$ in terms of 
the position vectors  ${\vec r}_j$ of the $j$th segments of a polymer consisting 
of $N$ segments for $j=1, 2, \ldots, N$, by 
\begin{equation}
\rho({\vec r}) = \sum_{j=1}^{N} \delta({\vec r}- {\vec r}_j) \, .  
\end{equation}

We call  $\langle \rho({\vec r}_1) \rho({\vec r}_2) \rangle$,  the statistical average of the product of the local densities at ${\vec r}_1$ and ${\vec r}_2$, the pair distribution function. \cite{Teraoka} 
Since the polymer system is homogeneous and isotropic in the ensemble average, we denote the pair distribution function $\langle \rho({\vec r}) \rho({\vec 0}) \rangle$ simply by $\langle \rho(r) \rho(0) \rangle$ with  $r=|{\vec r}|$. 
It is related to the radial distribution function of polymer segments, $g(r)$, as  
$g(r)= \langle \rho(r) \rho(0) \rangle/\rho^2$.

Suppose that we have a polymer of $N$ segments in a region of volume $V$.  
We can regard the local segment density at the origin normalized by the average $\rho= N/V$,  
i.e. $\rho({\vec 0})/\rho= \rho({\vec 0})/(N/V)$, as the expectation value for the number of a given segment to be found in a unit volume at the origin when we take the ensemble average. 
Intuitively, it corresponds to  the probability of finding a segment at the origin per volume. 
Thus, the average number of segments per volume located at ${\vec r}$ when there is already a segment at the origin ${\vec r}={\vec 0}$ is given by \cite{Teraoka} 
\begin{equation} 
\langle \rho({\vec r}) \rho({\vec 0}) \rangle/\rho \, .  
\end{equation}

\subsection{Exact expression of the pair distribution function of a Gaussian ring polymer}

Let us consider a Gaussian linear chain of $N$ steps with bond length $b$.  
For a given ${\vec r}_2$ the probability density for ${\vec r}_1$ is given by 
\begin{equation}
G({\vec r}_1, {\vec r}_2; N) = \left( 2 \pi N b^2/3 \right)^{-3/2} \exp\left(- \frac {\left( {\vec r}_1-{\vec r}_2 \right)^2} {2Nb^2/3} \right) \, . 
\label{eq:ete-gauss}
\end{equation}

For a Gaussian linear chain of $N$ steps 
the pair distribution function is given by   \cite{Teraoka} 
\begin{eqnarray}
& & \langle \rho({\vec r}) \rho(0) \rangle/\rho  =  \frac 1 N \int_{0}^{N} dn 
\int  _0^N G({\vec r}, {\vec 0}; |n-n^{'}| ) dn^{'} 
\nonumber \\ 
& & \quad = \frac 6 {\pi^{3/2} b^2 r} \int_{\sqrt{u}}^{\infty} \left(1- u /x^2 \right) \exp(-x^2) dx    
\end{eqnarray}
where $u$ is given by $u=3 r^2/(2Nb^2)$. For small $r$ we have
\begin{equation}
\langle \rho(r) \rho(0) \rangle/\rho \approx 1/r  
\label{eq:asymp-linear-small}
\end{equation} 
and for large $r$ we have with $R_G^2 = Nb^2/6$ 
\begin{equation}
\langle \rho(r) \rho(0) \rangle/\rho \approx \frac 1 {r^4} \exp(-r^2/4R_G^2) \, .   
\label{eq:asymp-linear-large}
\end{equation} 

For the Gaussian ring chain of $N$ steps  we formulate the pair distribution function in terms of the probability densities of the two random walks with $n$ and $N-n$ steps connecting the origin and ${\vec r}$ as follows.    
\begin{eqnarray}
& & \langle \rho({\vec r}) \rho({\vec 0}) \rangle /\rho \nonumber \\ 
& &  = \frac 1 N \int_0^{N}dn \int_0^{N} dn^{'} 
G({\vec r}, {\vec 0}; |n-n^{'}|)  G({\vec 0}, {\vec r}; N- |n-n^{'}|)  \nonumber \\ 
& & \quad \times \left(G({\vec 0}, {\vec 0}; N)  \right)^{-1} \, . 
\label{eq:def-ring}
\end{eqnarray} 
Here we remark that the normalization factor in (\ref{eq:def-ring}) 
is derived as follows.   
\begin{eqnarray}
G({\vec 0},{\vec 0}; N) & = &  
\int_V d^3{\vec r} \, \left( 
{\frac 1 N} \int_0^{N} dn  G({\vec 0}, {\vec r}; n) G({\vec r}, {\vec 0}; N-n) \right)  
\, . \nonumber \\
\end{eqnarray}
We can show that  eq. (\ref{eq:def-ring}) is expressed  as  
\begin{eqnarray}
& & \langle \rho({\vec r}) \rho({\vec 0}) \rangle /\rho 
\nonumber \\ 
& & \, \, = \int_{0}^{N} G({\vec r}, {\vec 0}; n) G({\vec 0}, {\vec r}; N-n) dn /G({\vec 0},{\vec 0}; N) \, . 
\label{eq:pair-polygon}
\end{eqnarray}
By calculating the integral (\ref{eq:pair-polygon}) with respect to variable $n$ 
we derive exactly an explicit 
expression of the pair distribution function of the Gaussian polygon of $N$ steps  
\begin{equation} 
\langle \rho(r) \rho({0}) \rangle /\rho 
=\frac{N} {2 \pi v^2} \, {\frac 1 r} \exp\left(- \frac {r^2} {v^2} \right) \,  ,  
\label{eq:pair-dist-polygon}
\end{equation} 
where $v^2$ is given by 
\begin{equation}
v^2 = {N b^2}/{6} \, .   
\end{equation}
Here we recall that for random polygons we have $R_G^2= N b^2/12$, and hence we have  
$v^2= 2 R_G^2$. 
The derivation of (\ref{eq:pair-polygon}) and (\ref{eq:pair-dist-polygon}) will be given in Appendix B. 

The exact expression (\ref{eq:pair-dist-polygon}) of the pair distribution function leads to the 
integral expression of the static structure factor derived by Casassa. \cite{Casassa}  
%

\subsection{Mean-square radius of gyration and hydrodynamic radius expressed with the distance distribution function}

We recall that the probability that the distance between two given segments  
is larger than $r$ and smaller than $r+dr$ is given by $F(r) dr $. 
It is related to the pair distribution function as 
\begin{equation}
N \, F(r) \, dr = {\frac 1 {\rho}} \langle \rho(r) \rho(0) \rangle \, 4 \pi r^2 dr  \, .  
\end{equation}

Let us denote by $r_{\rm RMS}$ the square root of the mean-square distance between two segments of a polymer. We define it by  
\begin{equation} 
\left( r_{\rm RMS}  \right)^2 = \int_{0}^{\infty} r^2 \, F(r)  \, dr \, . \label{eq:RMS}
\end{equation}
In terms of $r_{\rm RMS}$ we define the normalized distance by 
\begin{equation} 
x=r/r_{\rm RMS} .   
\end{equation} 
We then define the probability distribution function $f(x)$ of the normalized distance $x$ by 
\begin{equation} 
F(r) \, dr =f(r/r_{\rm RMS}) \, dr/r_{\rm RMS} \, . 
\end{equation} 
That is, we have assumed $F(r) dr = f(x) dx$.  We also call $f(x)$ the normalized distance  distribution, briefly.

It follows from (\ref{eq:RMS}) that not only the 0th moment but also the second moment of the normalized distance distribution is given by 1.  
The mean-square radius of gyration is expressed in terms of the normalized distance distribution as follows.  
\begin{eqnarray}
R_G^2 & = & {\frac 1 2} \,  \int_{0}^{\infty} r^2 \, F(r)  \, dr     \nonumber \\   
& = & {\frac 1 2} \,  r_{\rm RMS}^2  \,  \int_{0}^{\infty} x^2 \, f(x)  \, dx  \, . 
\label{eq:Rg}
\end{eqnarray}
Since the second moment of $f(x)$ is given by 1, we have 
\begin{equation}
R_G^2 = r_{\rm RMS}^2 /2\, . 
\end{equation}
We have $R_G= r_{\rm RMS}/\sqrt{2}$.  
The hydrodynamic radius is expressed in terms of the distance distribution function as  
\begin{eqnarray}
\frac 1 {R_H} & = &  \int_{0}^{\infty} {\frac 1 r} \,  F(r) \, dr  \nonumber \\ 
& = & {\frac 1 {r_{\rm RMS}}} \, \int_{0}^{\infty} x^{-1} f(x) dx \, .      
\end{eqnarray}
That is, the hydrodynamic radius $R_H$ is given by the harmonic mean of the distance between two segments. of the polymer.  
We therefore have the ratio of the gyration radius to the hydrodynamic radius as 
\begin{equation} 
R_G/R_H = {\frac 1 {\sqrt{2}}} \,  \int_{0}^{\infty} x^{-1} f(x) dx \, .    \label{eq:ratioRGRH}
\end{equation}

\subsection{Ideal topological polymers}

\begin{figure}[ht]
 \center
\includegraphics[clip,width=8.0cm]{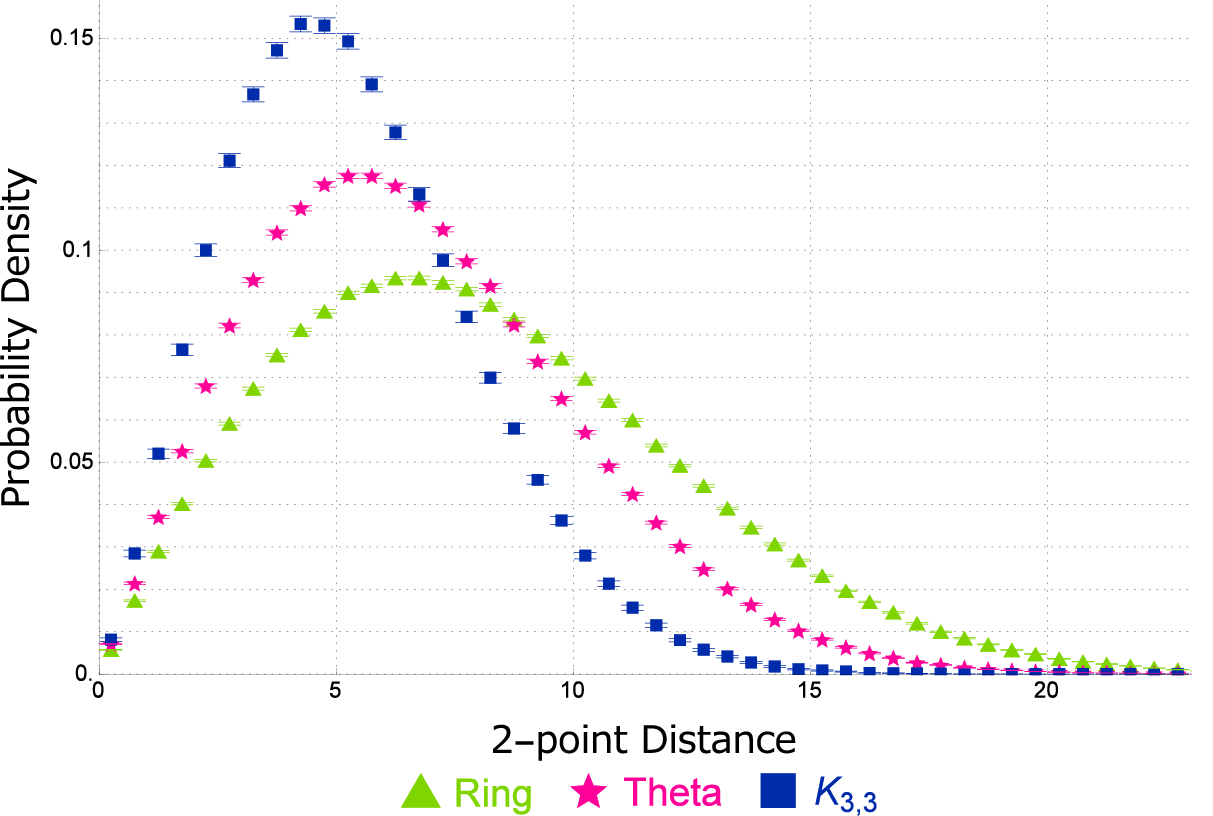} 
%
 \caption{Distance distribution function $F(r)$ of the distance $r$ between two segments in an ideal topological polymer of $N=500$ with graph $F$ for three graphs $F$: a complete bipartite graph $K_{3,3}$, a  theta curve, and a ring.  
Data points for the complete bipartite graph $K_{3,3}$,  theta curve and ring  correspond to filled squares, filled stars and filled triangles, respectively. 
 }
\label{fig:2pt-Q}
\end{figure}

The data points of the distance distribution function $F(r)$ for an ideal  topological polymer of $N=500$ segments with graph $F$ are plotted in Fig. \ref{fig:2pt-Q} against the distance $r$ between two segments of the polymer. Here we consider three graphs $F$, a ring,  a theta curve, and a complete bipartite graph $K_{3,3}$.  
The distance distribution function of a complete bipartite graph $K_{3,3}$ polymer has the tallest peak height among the ideal topological polymers of the three graphs.  That of a $\theta$-shaped polymer has the next highest, and that of a ring polymer the lowest peak height. Similarly, the peak positions of the distance distribution functions for a complete bipartite graph $K_{3,3}$ polymer, a $\theta$-shaped polymer, and a ring polymer  are given by the smallest one, the second smallest one, and the largest one, respectively.  
Accordingly, the distance distribution function of  
a complete bipartite graph $K_{3,3}$ polymer  has the narrowest width among the ideal topological polymers of the three graphs.

\begin{figure}[ht]
 \center
\includegraphics[clip,width=8.0cm]{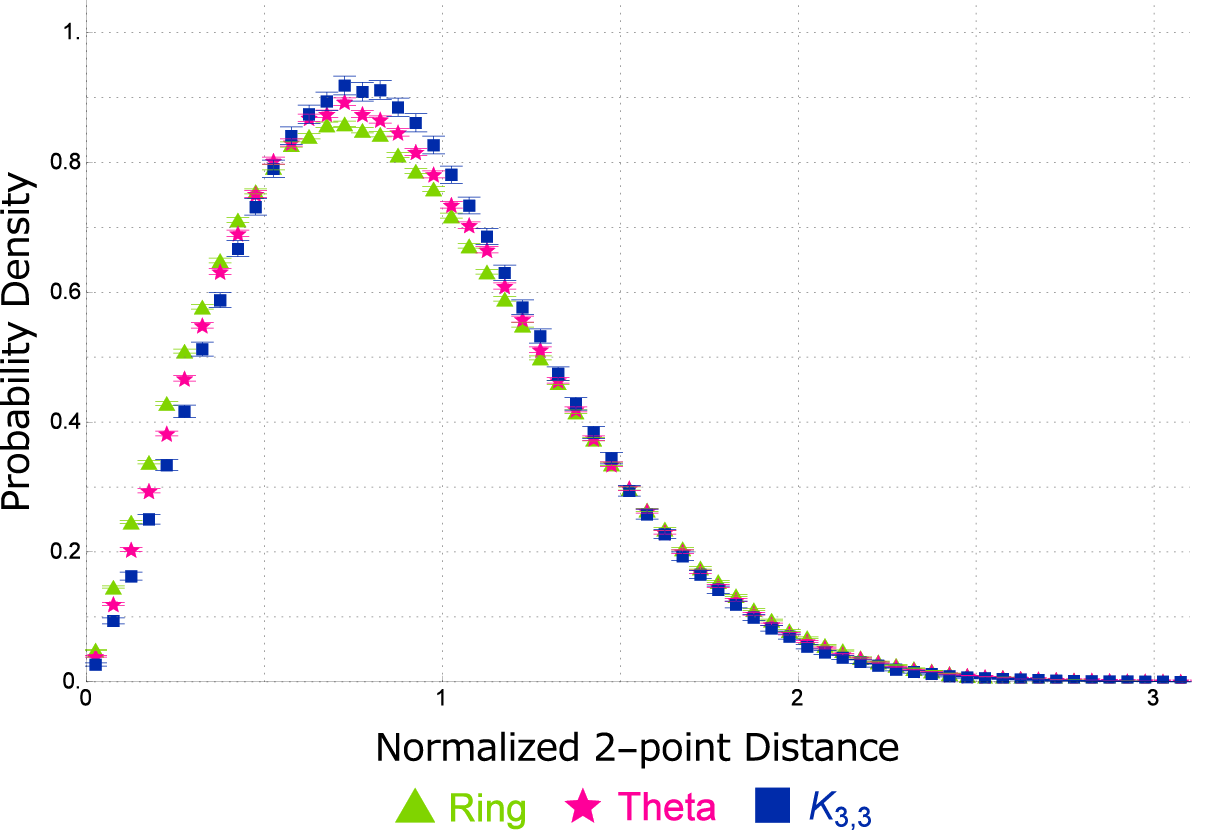} 
%
 \caption{Probability distribution function $f(x)$ of the normalized distance $x$ between any given pair of  segments in an ideal topological polymer of $N=500$ with graph $F$. Here the distance between two segments is normalized by the gyration radius of the topological polymer $R_g(F)$.}
\label{fig:2pt-Qnormalized}
\end{figure}

The different shapes of the plots depicted in Fig. \ref{fig:2pt-Q} for the three graphs are mainly due to the different values of  the root mean square  $r_{\rm RMS}$ for the three graphs.  In Fig. \ref{fig:2pt-Qnormalized} the data points of the normalized distance distribution $f(x)$ are plotted against the normalized distance $x=r/r_{\rm RMS}$. The three curves are almost overlapping 
and only slightly different from each other. However, such small different behavior may lead to quite different values of the ratio of the gyration radius to the hydrodynamic radius, $R_G/R_H$.

For the Gaussian ring chain, an exact expression of the pair distribution function and therefore the distance distribution function is obtained. The root mean-square distance $r_{\rm RMS}$ is given exactly by 
\begin{equation}
r_{\rm RMS}^2 = N b^2/6 .  
\end{equation}
In eq. (\ref{eq:pair-dist-polygon}) we have $4 r^2/v^2=r^2/2R_G^2 = (r/r_{\rm RMS} )^2$. We therefore have the exact expression of the normalized distance distribution 
\begin{equation} 
f(x) = 2 x \exp\left( -x^2 \right) \, . \label{eq:exact-f} 
\end{equation}
Here we recall that the normalized distance is given by $x=r/r_{\rm RMS}$.  

\begin{table}[htbp] 
\begin{tabular}{c|cccc} 
graph & $C$  & $\theta$ & $\delta$ & $\chi^2$/DF \\
\hline 
ring & $2.0028 \pm 0.0025$ & $1.0044 \pm 0.0020$ & $2.0011 \pm 0.0017$ & 0.96 \\ 
theta & $2.0796 \pm 0.0056$ & $1.0929 \pm 0.0043$ & $2.0661 \pm 0.0038$ & 2.39 \\    
$K_{3,3}$ & $2.1799 \pm 0.0092$ & $1.2096 \pm 0.0070$ & $2.1478 \pm 0.0060$ & 0.97 \\
\hline 
\end{tabular} 
\caption{Best estimate of the parameters in (\ref{eq:fit-2pt}) for the normalized distance distribution $f(x)$ for an ideal topological polymers of $N$ segments with graph $F$. Here $N=500$. Left-hand sides of (\ref{eq:con0th}) and (\ref{eq:con2nd}) are given by 0.9916 and 0.9985, respectively, for $\theta$-shaped polymers; by 0.9988 and 0.9984 for complete bipartite $K_{3,3}$ graph polymers, respectively.}
\label{tab:2pt} 
\end{table}

We shall numerically show that the normalized distance distribution $f(x)$ of an ideal  topological  polymer with graph $F$ is well approximated by the following 
\begin{equation} 
f(x) = C x^{\theta} \exp\left(- x^{\delta} \right) \label{eq:fit-2pt} \, .  
\end{equation} 
Here the three parameters satisfy the two constraints 
that the 0th and 2nd moments of $f(x)$  are given by 1   
\begin{eqnarray} 
{\frac C {\delta}} \Gamma((\theta+1)/\delta)  & = & 1 \, , \label{eq:con0th} \\  
{\frac C {\delta}} \Gamma((\theta+3)/\delta)  & = & 1 \, . \label{eq:con2nd} 
\end{eqnarray} 
For the Gaussian ring chain case, we have $C=2$, $\theta=1$ and $\delta=2$, exactly. It is clear that they satisfy the constraints (\ref{eq:con0th}) and (\ref{eq:con2nd}).

We now argue that formula (\ref{eq:fit-2pt}) gives good fitted curves to the data. 
The best estimates of the parameters $C$, $\theta$ and $\delta$ of eq. (\ref{eq:fit-2pt}) are listed in Table \ref{tab:2pt}. The $\chi^2$ values per DF are rather small for the three graphs: a ring, a $\theta$-shaped curve and a complete bipartite $K_{3,3}$ graph. 
For ring polymers the best estimates are close to the exact values: $C=2$, $\theta=1$ and $\delta=2$. 
By putting the best estimates in Table \ref{tab:2pt} we evaluate the left hand-sides of (\ref{eq:con0th}) and (\ref{eq:con2nd}). They hold within errors for $\theta$-shaped polymers and complete bipartite $K_{3,3}$ graph polymers as shown in the caption of Table \ref{tab:2pt}.  
We thus conclude that formula (\ref{eq:fit-2pt}) gives good fitted curves to the data.

Substituting (\ref{eq:fit-2pt}) to the integral of (\ref{eq:ratioRGRH}) we have 
\begin{equation}
R_G/R_H = {\frac 1 {\sqrt{2}}} {\frac C {\delta}} \Gamma({\theta}/{\delta}) \, . 
\label{eq:inverse}
\end{equation}  
For the Gaussian ring polymer, the ratio $ R_G/R_H$ is exactly given by 
\begin{equation}
R_G/R_H = \sqrt{ \frac {\pi} 2}  \, .   \label{eq:ratio-ring} 
\end{equation}
The numerical value $\sqrt{\pi/2} \approx 1.253314$ is rather close to the estimate obtained by the quaternion method presented in Table \ref{tab:ratioN500}. 

Putting into (\ref{eq:inverse}) the best estimates of the parameters of (\ref{eq:fit-2pt}) given in Table \ref{tab:2pt} we evaluate the ratio $R_G/R_H$ as follows: 1.24966 for ring polymers, 1.19415 for $\theta$-shaped polymers, and 1.13402 for complete bipartite $K_{3,3}$ graph polymers. They are all consistent with the estimates of $R_G/R_H$ listed in Table \ref{tab:ratioN500}. 

%
\begin{figure}[ht]
 \center
\includegraphics[clip,width=8.0cm]{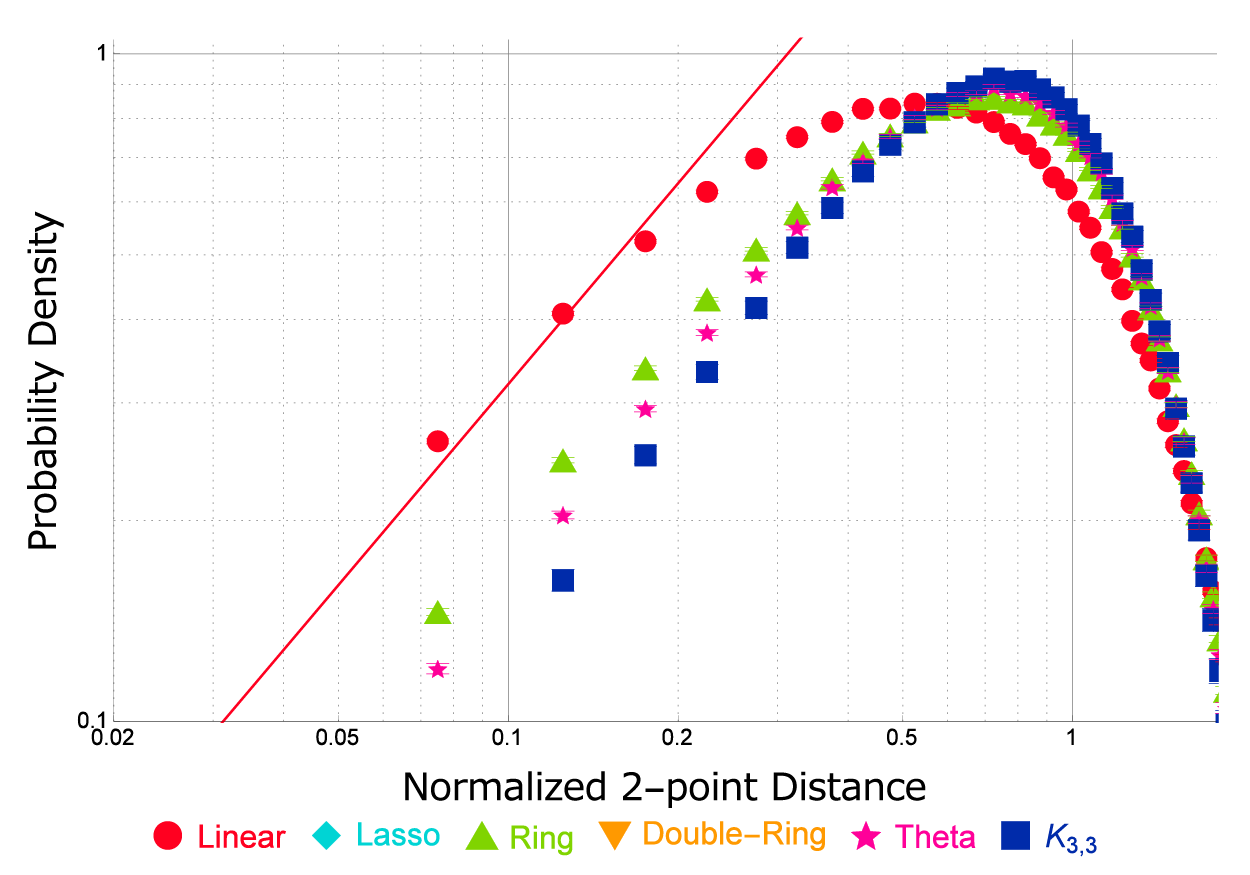}
%
\caption{Double-logarithmic plot of the normalized distance distribution $f(x)$ in an ideal topological polymer of $N=500$ with graph $F$. 
The data points for a complete bipartite graph $K_{3,3}$ polymer, a $\theta$-shaped polymer,  a ring polymer and a linear polymer correspond to  filled squares, filled stars, filled triangles, and filled circles, respectively.
}
\label{fig:2pt-Qnormalized-DL}
\end{figure}

In order to investigate the asymptotic behavior the double-logarithmic plots of the normalized distance distribution $f(x)$  for ideal topological polymers of $N=500$ are given for the four graphs $F$ in Fig. \ref{fig:2pt-Qnormalized-DL}. The short-distance behavior of the normalized distance distribution of a linear chain is consistent with the analytical result (\ref{eq:asymp-linear-small}) such as $f(x) \approx x$ for $x \ll 1$. The large-distance behavior of a normalized distance distribution is consistent with the analytic result (\ref{eq:asymp-linear-large}) such as $f(x) \approx \exp(-x ^2/2)$ for $x \gg 1$. For the Gaussian ring chain we have both the short-distance and the large-distance behavior in the exact expression of the normalized distance distribution: $f(x) =2 x \exp(-x^2)$.

\subsection{Real topological polymers}

\begin{figure}[ht] 
 \center
\includegraphics[clip,width=9.0cm]{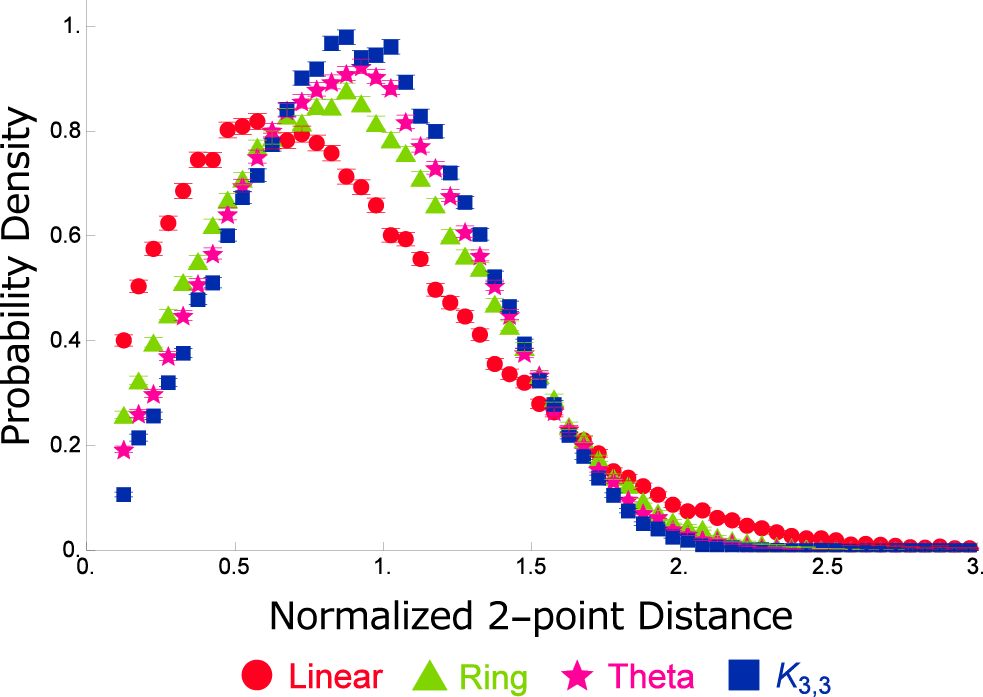}
%
\caption{Probability distribution function $f(x)$ of the normalized distance $x$ 
between two segments in a real topological polymer (the Kremer-Grest model) of $N=300$ with graph $F$. Here, graphs $F$ correspond to a linear, a ring, a theta curve and a complete bipartite graph $K_{3,3}$ polymer, respectively. }
\label{fig:2pt-realLL}
\end{figure}

The estimates of the probability distribution function $f(x)$ of the normalized distance $x$  between two segments in a real topological polymer of $N=300$ with graph $F$ are plotted in Fig. \ref{fig:2pt-realLL} for four graphs: a linear, a ring, a theta curve, and a complete bipartite  $K_{3,3}$ graph. 
In Fig. \ref{fig:2pt-realLL} the normalized distance distribution of complete bipartite $K_{3,3}$ graph polymers  has the highest peak,  that of  theta curve polymers the second highest peak,and that of ring polymers the third highest peak.  Accordingly, the normalized distance distribution of complete bipartite graph $K_{3,3}$ polymers has the narrowest width among the real  topological polymers of the three graphs: a ring, a theta curve, and a complete bipartite  $K_{3,3}$ graph.

\begin{figure}[ht] 
\center
\includegraphics[clip,width=8.0cm]{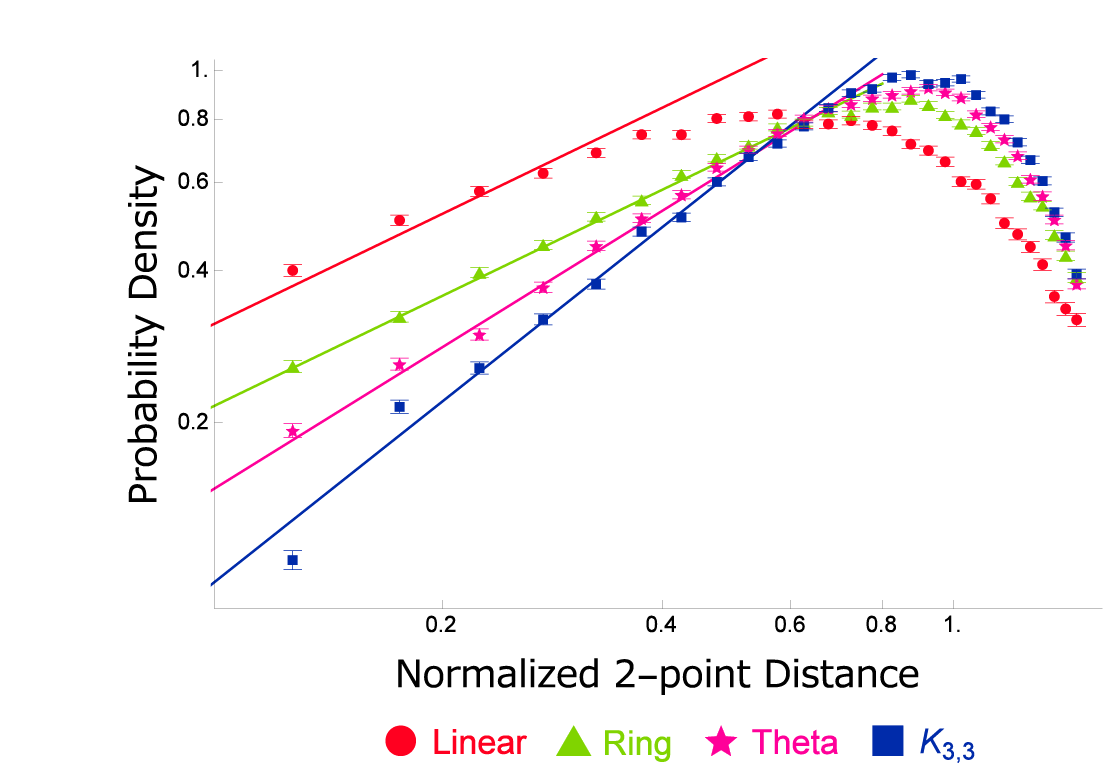} 
%
\caption{Double logarithmic plot of the probability distribution function $f(x)$ of the normalized distance $x$ between two segments in a real topological polymer (the Kremer-Grest model) of $N=300$ with graph $F$. The fitted lines are given by 
$1.6 \, x^{0.7}$, $1.1 \, x^{0.7}$, $1.2 \, x^{0.9}$  and $1.4 \, x^{1.15}$  for a linear, a ring,  a theta curve, and a complete bipartite graph $K_{3,3}$ polymer, respectively. }
\label{fig:2pt-normalizedDL}
\end{figure}

The probability distribution function $f(x)$ of the normalized distance $x$ between two segments in a real topological polymer (the Kremer-Grest model) of $N=300$ with graph $F$ is plotted in the double logarithmic scale in Fig. \ref{fig:2pt-normalizedDL} for each of the four graphs such as linear, ring, $\theta$-shaped and complete bipartite $K_{3,3}$ graphs. 

The enhancement of the short-distance correlation is characterized by the different values of 
the exponent for the short-distance power-law behavior. It is given by $0.7$ for both a linear and a ring polymer, while $0.9$ and $1.15$ for a theta curve polymer and a complete bipartite graph $K_{3,3}$ polymer, respectively, as shown in  Fig. \ref{fig:2pt-normalizedDL}.

We suggest that the short-distance exponent of a topological polymer is given by the same value for different numbers of segments $N$. For $N=400$ we have plotted the normalized distance distribution function in a real topological polymer (the Kremer-Grest model) for the four different graphs.  We have the same fitted lines for the power-law behavior for the four graphs. That is, the exponent is given by $0.7$ for both a linear and a ring polymer, while it is given by $0.9$ and $1.15$  for a theta curve polymer and a complete bipartite graph $K_{3,3}$ polymer, respectively.

For star polymers the Kratky plot of the static structure factor is calculated for real chains. \cite{TKC02} It seems, however, that the large-$q$ behavior is similar to that of an open chain.

%
%
\section{Concluding remarks} 

For various polymers with different topological structures we have numerically evaluated the mean-square radius of gyration and the hydrodynamic radius systematically through simulation. We evaluated the two quantities both for ideal and real chain models and show that the ratios of the quantities among different topological types do not depend on the existence of excluded volume if the topological polymers have only up to trivalent vertices, as far as the polymers investigated. 

We have shown that the quaternion method for generating ideal topological polymers is practically quite useful for evaluating physical quantities for the topological polymers.  In fact, we can estimate at least approximately the values of $R_G^2$ or $R_H$ for real topological polymers by making use of the ratios among the corresponding ideal ones. Furthermore, the quaternion algorithm is quite fast. The computational time is proportional to the number of segments.  

For various topological polymers we have evaluated the ratio of the gyration radius to the hydrodynamic radius, $R_G/R_H$, which we expect to be universal from the viewpoint of renormalization group.  We have analytically shown that the ratio $R_G/R_H$ of a topological polymer is characterized with the variance of the distance distribution function, i.e. the probability distribution function of the distance between two given segments of the polymer. If the variance decreases, then the ratio $R_G/R_H$ also decreases.

We have shown that the short-range intrachain correlation is much enhanced for real topological polymers expressed by complex graphs. We suggest that the correlation hole becomes large for topological polymers since  the excluded-volume effect is enhanced due to the constraints derived from the complex structure of the polymers. The short-range correlation is characterized by the power-law behavior, and the value of the exponent increases as the graph becomes more complex.  
It would therefore be interesting to observe  the enhanced short-range correlation of topological polymers in the large-$q$ region of the static structure factor through scattering experiments.   

\section*{Acknowledgements} 

We would like to thank Profs. K. Shimokawa and Y. Tezuka for helpful discussion on topological polymers.  The present research is partially supported by the Grant-in-Aid for Scientific Research No. 26310206. 

\appendix

\setcounter{equation}{0} 
\renewcommand{\theequation}{A.\arabic{equation}}

\section{Diffusion coefficients of topological polymers }
\begin{figure}[ht]
\center
\includegraphics[clip,width=8.0cm]{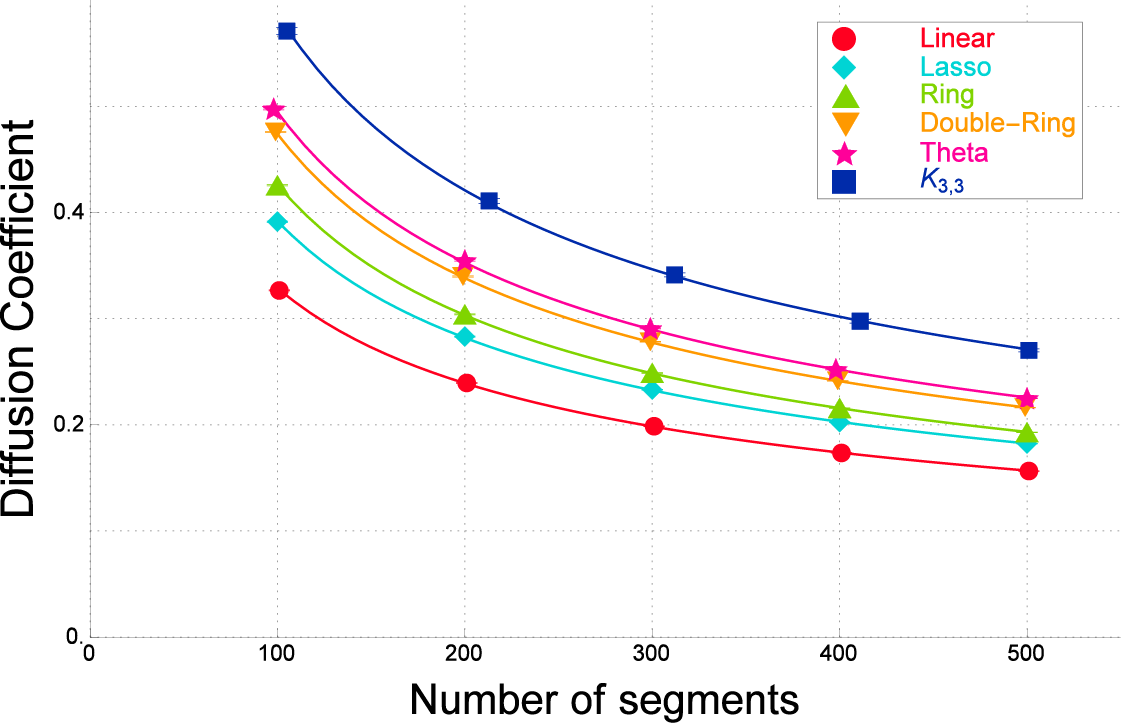}
%
\caption{Diffusion coefficient $D(F)$ versus the number of segments $N$ for the ideal topological polymer with graph $F$ evaluated  by the quaternion method. }
\label{fig:QD}
\end{figure}

We define the diffusion coefficient $D(F)$ by the inverse of the hydrodynamic radius $R_H(F)$ 
for a topological polymer of graph $F$; $D(F)=1/R_H(F)$.  

In Fig. \ref{fig:QD} the estimates of the diffusion coefficient $D(F)$ for ideal topological polymers 
with several different graphs $F$ are plotted against the number of segments $N$. 
The curves fitted to the data points are given by applying the following formula  
\begin{equation} 
D(F) = b_1 N^{-\nu} \, .  \label{eq:fit-QD}
\end{equation}
The best estimates of the parameters $b_1$ and $\nu$ are listed in  
Table \ref{tab:QD}.

\begin{table}[htbp] 
\begin{tabular}{c|ccc} 
graph $F$ & $b_1$  &  $\nu$  & $\chi^2$/DF \\  \hline 
linear & $2.735 \pm 0.039$ & $0.4599 \pm 0.0025$ & $4.37$  \\ 
tadpole (lasso) & $3.500 \pm 0.042$ & $0.4754 \pm 0.0021$ & $3.57$  \\ 
ring & $4.098 \pm 0.038$ & $0.4914 \pm 0.0017$ & $2.21$  \\ 
double-ring & $4.480 \pm 0.033$ & $0.4877 \pm 0.0013$ & $1.51$  \\ 
$\theta$-shaped & $4.670 \pm 0.092$ & $0.4875 \pm 0.0035$ & $6.3$  \\ 
complete $K_{3,3}$ & $5.350 \pm 0.130$ & $0.4797 \pm 0.0042$ & $0.87$  \\ \hline 
\end{tabular} 
\caption{Best estimates of the parameters of  (\ref{eq:fit-QD}) 
to the diffusion coefficients of ideal topological polymers versus $N$ for various graphs $F$ and the $\chi^2$ values per DF. } \label{tab:QD}
\end{table}

\begin{figure}[ht]
\center
\includegraphics[clip,width=8.0cm]{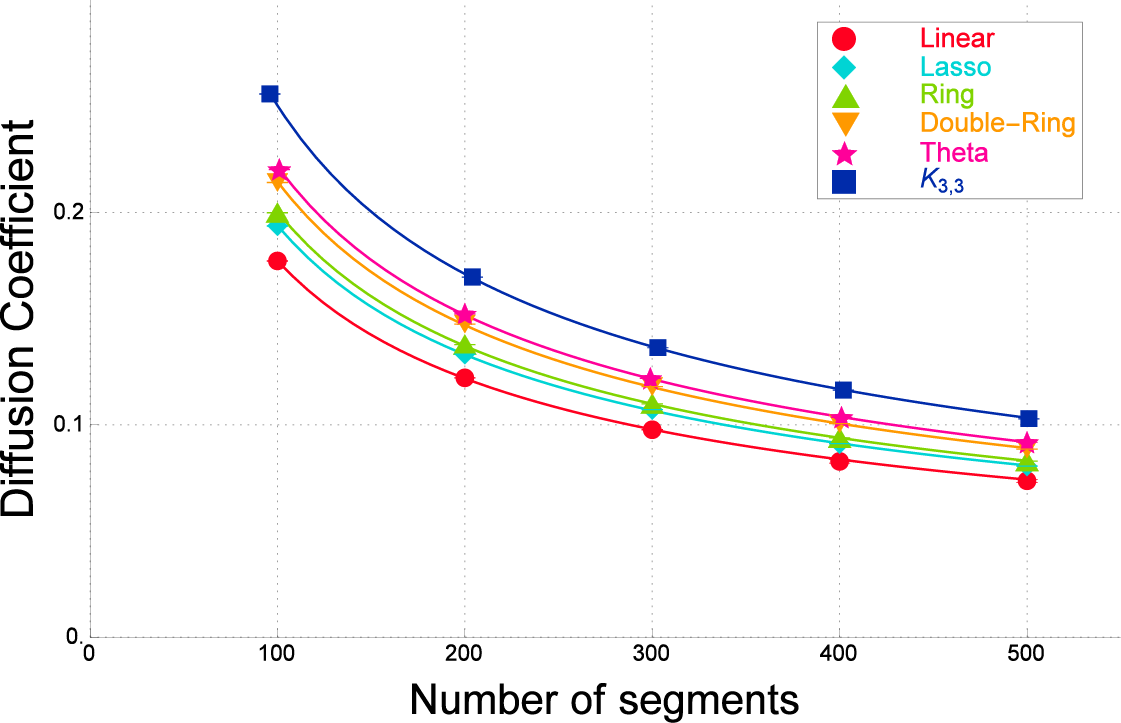}
%
 \caption{Diffusion coefficient $D(F)$ versus the number of segments $N$ for topological polymers of graph $F$ with excluded volume  (the Kremer-Grest model).}
\label{fig:LD}
\end{figure}

In Fig. \ref{fig:LD} the estimates of the diffusion coefficient $D(F)$ for real topological polymers with several different graphs $F$ are plotted against the number of segments $N$. The curves fitted to the data points are given by (\ref{eq:fit-QD}). The best estimates of the parameters $b_1$ and $\nu$ are listed in  Table \ref{tab:LD}.

\begin{table}[htbp] \begin{tabular}{c|ccc} 
graph & $b_1$  & $\nu$ & $\chi^2$/DF \\ \hline 
linear & $2.147 \pm 0.028$ & $0.5415 \pm 0.0025$ & $2.13$  \\ 
tadpole (lasso) & $2.372 \pm 0.033$ & $0.5437 \pm 0.0025$ & $6.93$  \\ 
ring & $2.489 \pm 0.035$ & $0.5471 \pm 0.0025$ & $9.51$  \\ 
double-ring & $2.661 \pm 0.043$ & $0.5467 \pm 0.003$ & $14.33$  \\ 
$\theta$-shaped & $2.764 \pm 0.027$ & $0.5478 \pm 0.0018$ & $7.54$  \\ 
complete $K_{3,3}$ & $3.154 \pm 0.032$ & $0.5501 \pm 0.0019$ & $13.13$  \\ \hline 
\end{tabular} 
\caption{Best estimates of the parameters in eq. (\ref{eq:fit-QD}) for the diffusion coefficients of topological polymers with excluded volume (the Kremer-Grest model) of $N$ segments for various graphs and the $\chi^2$ values per DF.} \label{tab:LD} 
\end{table}

\section{Derivation of the pair distribution function of a Gaussian ring chain}

Let us formulate the double integral in (\ref{eq:def-ring}) into a single integral 
(\ref{eq:pair-polygon}). We divide the square region $n$ and $n^{'}$ for $0 \le n, n^{'} \le N$ into the region of $n> n^{'}$ and that of $n < n^{'}$.  In the former region we define $m$ by $m=n-n^{'}$, and we have
\begin{eqnarray}
& & \int_0^N dn \int_0^n dn^{'} G({\vec r}, {\vec 0}; |n-n^{'}|) G({\vec 0}, {\vec r}; N-|n-n^{'}|)
\nonumber \\ 
& & = \int_0^N dn \int_0^n dm 
G({\vec r}, {\vec 0}; m) G({\vec 0}, {\vec r}; N-m) \, . \label{eq:gt}
\end{eqnarray}
In the latter we define $m$ by $m=N-(n^{'}-n)$, and we have   
\begin{eqnarray}
& & \int_0^N dn \int_n^N dn^{'} 
G({\vec r}, {\vec 0}; |n-n^{'}|) G({\vec 0}, {\vec r}; N-|n-n^{'}|) \nonumber  \\ 
& & = \int_0^N dn \int_n^N dm \, 
G({\vec r}, {\vec 0}; N-m) G({\vec 0}, {\vec r}; m)  \, . \label{eq:lt}
\end{eqnarray}
Since $G({\vec 0}, {\vec r}; m)=G({\vec r}, {\vec 0}; m)$ and $G({\vec r}, {\vec 0}; N-m) = G({\vec 0}, {\vec r}; N-m)$ after taking the sum of (\ref{eq:gt}) and (\ref{eq:lt}) we derive the single integral (\ref{eq:pair-polygon}). 

Making use of the probability density of Gaussian chains we have 
\begin{eqnarray} 
& & G({\vec r}, {\vec 0}; m) G({\vec 0}, {\vec r}; N-m)/G(0,0; N) \nonumber \\ 
& & = \left( {\frac {2 \pi n(N-n) b^2/3} {N}}  \right)^{-3/2}   \exp\left(- \frac {r^2} { (2n(N-n)/N) (b^2/3 )}  \right) \, . \nonumber \\
& & = G({\vec r}, {\vec 0}; n(N-n)/N) 
\end{eqnarray} 
Due to the symmetry of a ring the integral of $n$ from 0 to $N/2$ and from $N/2$ to $N$ give the same result, and hence we have 
\begin{equation}
\int_{0}^{N} G({\vec r}, {\vec 0}; n(N-n)/N) \, dn 
= 2 \int_{0}^{N/2} G({\vec r}, {\vec 0}; n(N-n)/N) \, dn \, . 
\end{equation}
We now define variable $z$ by 
\begin{equation} 
1/z = (2n(N-n)/N) (b^2/3)
\end{equation}
In terms of variable $z$ parameter $n$ is expressed as 
\begin{equation}
n = {\frac N 2}  \left( 1 \pm \sqrt{1- {\frac 1 {v^2 z}} }  \right) \, ,  
\end{equation}
where the minus symbol is for $0 \le n \le N/2$ and the plus symbol for $N/2 \le n \le N$. We then introduce variable $t$ by $t=z- 1/v^2$. We thus have  
\begin{eqnarray}
& & 2 \int_{0}^{N/2} G({\vec r}, {\vec 0}; n(N-n)/N) \, dn  \nonumber \\  
& & = {\frac {N} {2 \pi^{3/2} v^2} } \int_{1/v^2}^{\infty} 
\frac {\exp\left( - r^2 z \right)} { \sqrt{z- 1/v^2} } \, dz  \nonumber \\ 
& = &  {\frac {N} {2 \pi^{3/2} v^2} }  \exp(- r^2/v^2)  
\int_{0}^{\infty} t^{-1/2} \exp\left( - r^2 t \right)  \, dt \nonumber \\  
& = & {\frac {N} {2 \pi v^2}} \, {\frac 1 r} \, \exp\left(- \frac { r^2} {v^2} \right) \, . 
\end{eqnarray} 


\end{document}